\documentclass[final,3p,times,twocolumn,authoryear]{elsarticle}

\usepackage{graphicx}
\usepackage{textgreek}
\usepackage{amsmath}
\usepackage{csquotes}
\usepackage{gensymb}
\usepackage{subcaption}  % For subfigures
\usepackage{textcomp}
\usepackage{xcolor} % Required for text color
\hfuzz=\maxdimen
\vfuzz=\maxdimen

\usepackage{amssymb}
\usepackage{amsthm}

 \usepackage{xcolor}

 \usepackage[citebordercolor=white]{hyperref}
 \hypersetup{colorlinks=true,linkcolor={blue},citecolor={blue},urlcolor={red}}

\journal{icarus}

\begin{document}

\begin{frontmatter}

\title{Study on the Venusian Atmospheric Thermal Structure: A Comparative Analysis between VEX/Akatsuki and Venus-GRAM/VCD} %% Article title

\author[label1,label2]{Soumyaneal Banerjee\corref{cor1}}
\author[label1]{R. K. Choudhary}
\author[label3]{Keshav R Tripathi} %% Author name
\author[label1]{K. M. Ambili}
\author[label4]{J. Oschlisniok}
\author[label4]{Silvia Tellmann}
\author[label5]{T. Imamura}
\author[label6]{H. Ando}

\affiliation[label1]{organization={Space Physics Laboratory},%Department and Organization
            addressline={VSSC},
            city={Thiruvananthapuram},
            postcode={695022}, 
            state={Kerala},
            country={India}}
\affiliation[label2]{organization={Research Centre},%Department and Organization
            addressline={University of Kerala},
            city={Thiruvananthapuram},
            postcode={695034}, 
            state={Kerala},
            country={India}}
\affiliation[label3]{organization={Physical Research Laboratory},%Department and Organization
            city={Ahmedabad},
            postcode={380009},
            state={Gujarat},
            country={India}}
\affiliation[label4]{organization={Rheinisches Institut fur Umweltforschung, Freies Institut fur Planetenforschung},%Department and Organization
            city={Cologne},
            postcode={50931},
            country={Germany}}
\affiliation[label5]{organization={Graduate School of Frontier Sciences, The University of Tokyo},%Department and Organization
            city={Kashiwa},
            postcode={277-8561},
            state={Chiba},
            country={Japan}}
\affiliation[label6]{organization={Faculty of Science, Kyoto Sangyo University},%Department and Organization
            city={Kyoto},
            postcode={603-8555},
            country={Japan}}

\cortext[cor1]{Corresponding author: soumyaneal08@gmail.com}

% \date{\small Accepted for publication in Icarus. \\ \vspace{0.2cm} \today}

%% Abstract
\begin{abstract}
We investigate the thermal structure of the Venusian middle atmosphere between 45 and 80 km using radio occultation (RO) measurements from Venus Express and Akatsuki spanning 2006--2024. Retrieved temperature profiles are compared with climatological predictions from the Venus-GRAM and the Venus Climate Database (VCD). Systematic deviations exceeding 10 K are observed between the RO temperatures and model climatologies, particularly at high latitudes in both hemispheres. The long-term dataset further reveals a possible decadal-scale temporal variability in temperatures across the low-to-mid latitude regions. This variability becomes less coherent at higher altitudes. A combination of post-stratification and bootstrap analysis on the temperature anomalies indicates that the observed temporal variability at low-to-mid and polar latitudes is not readily explained by the latitudinal sampling bias quantified using the adopted analysis, while the trends at mid-to-high latitudes are affected by sparse and uneven RO sounding. Sensitivity tests using modified cloud albedo inputs in VCD simulations show that adjusting cloud radiative forcing partially reconciles the discrepancies, especially at the lower altitudes, below 60 km. However, above the cloud top, at the lower latitudes, the divergence from the observations increases significantly, thereby failing to provide a general reconciliation between VCD and RO. These results highlight the need for updated empirical climatologies incorporating recent RO measurements and for improved physical parameterizations in Venus general circulation models to better capture the global variability in the planet’s middle atmosphere.
\end{abstract}

%% Keywords

\end{frontmatter}

%\linenumbers


\section{Introduction}
\label{Intro}
%% Labels are used to cross-reference an item using \ref command.

The Venusian atmosphere has been extensively studied since the advent of the space age in the 1960s. Spearheaded by the numerous Venera and VEGA missions of the USSR and NASA missions such as Pioneer Venus and Magellan in the 1970s–1990s, the legacy continued with ESA's Venus Express and JAXA's Akatsuki mission \citep{Avduevsky_1970, Colin_1980, Saunders_1991, Svedhem_2007, Nakamura_2011}. In the near future, several new Venus missions with exciting prospects are planned, expected to launch toward the end of this decade and continue into the 2030s \citep{Garvin_2022, StraumeLindner_2022}. In addition to spacecraft missions, ground-based telescopic studies and modeling efforts have provided fundamental theoretical and experimental support, adding necessary redundancy to our primary knowledge base \citep{Meadows_1996, Lebonnois_2016}.

Through these studies, numerous fascinating and complex aspects of the Venusian atmosphere have emerged. These include extreme surface temperatures and pressures ($\sim$737 K, $\sim$92 bar), atmospheric super-rotation, optically thick global sulfuric acid clouds, warm polar regions above cloud-top altitudes ($\sim$65 km), complex cloud sulfur chemistry, and wave phenomena across various temporal and spatial scales (such as gravity and planetary waves). A comprehensive compilation and summary of these findings can be found in \citep{Venus1, Venus2, Venus3}.The thermal structure of the Venusian atmosphere is of particular interest to researchers. In the lower and middle atmosphere, vertical and horizontal temperature gradients drive atmospheric circulation, impacting wave dynamics. Temperature and pressure changes modulate the extent of cloud cover and thus impact crucial microphysical processes within the clouds, also affecting the radiative balance of the atmosphere. Understanding the thermal structure and its potential evolution over short and long timescales serves as an effective tool for comparative planetological studies, including those related to the theorized runaway greenhouse effect on Venus and Earth.

Ground-based telescopic microwave studies in the 1950s provided the first indication of high temperatures on the surface and in the lower atmospheric regions of Venus \citep{Mayer_1958}. This was confirmed by the Venera and Pioneer Venus (PV) landers in the 1960s and 1970s \citep{SEIFF_2022}. Radio Occultation (RO) subsequently provided a more comprehensive description of temperatures across the 40–90 km altitude range \citep{Kliore_1982}. Coupled with the PV Orbiter Infrared Radiometer (OIR), RO yielded vertical profiles of the southern hemispheric regions and provided the first evidence for the cold-collar feature -- a region of low temperatures centered around 65° latitude in the 60–70 km altitude range, surrounded by warm polar temperatures \citep{SCHOFIELD_1983}.

The datasets from the landers and orbiters of the Venera and PV missions formed the basis for the Venus International Reference Atmosphere (VIRA) \citep{Seiff_1985}, a semi-empirical model that has served as the reference for subsequent Venus missions. More details about early understanding of the thermal structure of Venus's atmosphere can be found in \citet{CRISP_2022}. Data from post-PV missions, including Venera 15 and 16, VEGA 1 and 2 landers and balloons, and Magellan RO, were incorporated with VIRA into an updated version, VIRA2 \citep{Moroz_1997}. The new datasets reaffirmed the averaged outputs of VIRA, although local time and latitudinal variabilities were also observed \citep{Zasova_2007}.

The Venus Express (VEX) and Akatsuki missions, which studied Venus from 2006 to 2014 and 2016 to 2024, respectively, have significantly enhanced our understanding of the Venusian atmosphere. \citet{Tellmann_2009} analyzed VEX radio occultation (RO) data and confirmed the existence of the cold collar in the northern hemisphere. Importantly, general agreement between PV Orbiter Radio Occultation (ORO) and VEX RO temperatures demonstrated their utility for studying the long-term stability of the middle and lower atmosphere. \citet{Ando2020} studied RO data from the VEX and Akatsuki missions and reported a low-stability layer in the polar regions down to an altitude of 42 km. They also compared RO temperatures with the VIRA and VIRA2 models and found good agreement, with a maximum discrepancy of less than 10 K. \citet{Ando_2025} combined RO datasets from VEX and Akatsuki to investigate the long-term evolution of the middle atmosphere's thermal structure. Their results revealed significant temporal variability exceeding 10 K, linked to changes in cloud UV albedo and zonal winds near the cloud top. These findings suggest a complex coupling between atmospheric dynamics, cloud microphysics, and photochemistry, although no definitive mechanism has yet been established. Their analysis, however, was limited to low-latitude regions, leaving potential temporal variability at higher latitudes largely unexplored.

Despite these advances, several key aspects of the Venusian atmospheric thermal structure remain insufficiently constrained, particularly regarding global variability and consistency with existing atmospheric models. The limited latitudinal coverage and the absence of an established physical mechanism linking thermal variability to cloud and dynamical processes highlight the need for a more comprehensive investigation. Thus, the thermal structure of the Venusian atmosphere forms the central focus of this study. The objective is to examine the global thermal structure between 45–80 km altitudes derived from VEX RO data and compare it with temperature profiles from the Venus-GRAM (VGRAM) and Venus Climate Database (VCD) models to assess their reproducibility and identify regions of disagreement. Additionally, the long-term temporal evolution of middle atmospheric temperatures is investigated using combined RO datasets from VEX and Akatsuki spanning 2006 to 2024. Low-to-mid and polar-latitude observations are analyzed and compared with model outputs, and the impact of varying cloud albedo conditions on VCD-derived temperatures is evaluated.

\section{Data and Methodology} \label{data}

\subsection{The RO Technique} \label{RO}

RO is a remote sensing technique that enables the sounding of planetary atmospheres at high vertical resolution and provides dense coverage in latitude and local time \citep{Eshleman_1973, Withers_2017}. It is especially valuable for studying the Venusian atmosphere due to its ability to penetrate the dense, opaque global sulfuric acid clouds, which otherwise make direct observations below the cloud-top altitudes ($\sim$65–70 km) extremely difficult. Radio science experiments have been an integral component of most Venus missions, whether flybys or orbiters \citep{Fjeldbo_1971, Hausler_2006, Imamura_2017}.

In a typical Venus RO experiment, an ultrastable oscillator onboard the spacecraft emits a radio signal (typically in the S-band or X-band) that traverses the Venusian atmosphere, undergoing phase shifts and amplitude attenuation before being recorded by a ground station on Earth. The phase shift in the received electromagnetic (EM) wave provides information on the bending angle along the ray path, which is inverted to derive the vertical refractive index profile ($\mu$) of the atmosphere via an Abel transform \citep{Fjeldbo_1971, Tripathi2022a, Tripathi2022b}. The atmospheric neutral number density ($n$) is calculated from the relation $\mu - 1 = K \cdot n$, where $K$ is the mean refractive volume of the Venusian atmosphere ($\sim 1.804 \times 10^{-29}\text{ m}^3$) \citep{Essen1951, Hinson1999}. Assuming a well-mixed atmosphere below 100 km dominated by CO$_2$ ($\sim$96.5\%) and N$_2$ ($\sim$3.5\%), the hydrostatic equation is integrated to derive the vertical temperature profile subject to an upper boundary condition \citep{lipa1979statistical, Tellmann_2009, Tripathi2022a}:

\begin{equation}
T(h) = \frac{\mu_{\text{top}}}{\mu(h)} \cdot T_{\text{top}} + \frac{\overline{m}}{k_B \cdot n(h)} \int_{h}^{h_{\text{top}}} n(h') \cdot g(h')    dh'
\label{eq:temperature_eqn}
\end{equation}

where $T(h)$ is the temperature at altitude $h$, $\overline{m}$ is the mean molecular mass of the neutral atmosphere, $k_B$ is the Boltzmann constant, and $g(h')$ is the altitude-dependent gravitational acceleration of Venus. The boundary condition assumes a constant temperature at the upper altitude limit (typically 90–100 km for Venus), below which hydrostatic equilibrium is considered. Because the influence of the chosen boundary temperature becomes negligible below $\sim$90 km \citep{Tellmann_2009}, the temperature at 100 km is set to 200 K for this study.

\subsection{The RO Datasets}
\label{RO_data}

RO data from Venus Express and Akatsuki have been used to derive the temperature profiles from 2006-2024. The VeRa payload onboard VEX gave data from July 2006 to March 2014. It conducted the experiments in one-way downlink mode at S-band ($\sim$2.3 GHz) and X-band ($\sim$8.4 GHz) frequencies and was received by ground stations of ESA and NASA (ESTRACK and DSN, respectively). Akatsuki’s RS payload was operational from March 2016 till April 2024. Similar to VEX, one-way downlink RO experiments were carried out in the single frequency band (X-band, $\sim$ 8.4 GHz) and received at the deep space network stations of JAXA and ISRO, and at the DLR Weilheim station in Germany. Ultra-stable oscillators onboard both the spacecraft served as the radio signal generators and ensured very low uncertainty in the emitted signal phase and the subsequently estimated temperatures. Both payloads had respective stabilities of the order of $10^{-12}$ over 100s and 1-1000s integration time, leading to measured temperature uncertainties between 0.1-1 K \citep{Tellmann_2009, Imamura_2011}.

\begin{figure*} [tbh]
    \centering
    \includegraphics[width=1.0\linewidth]{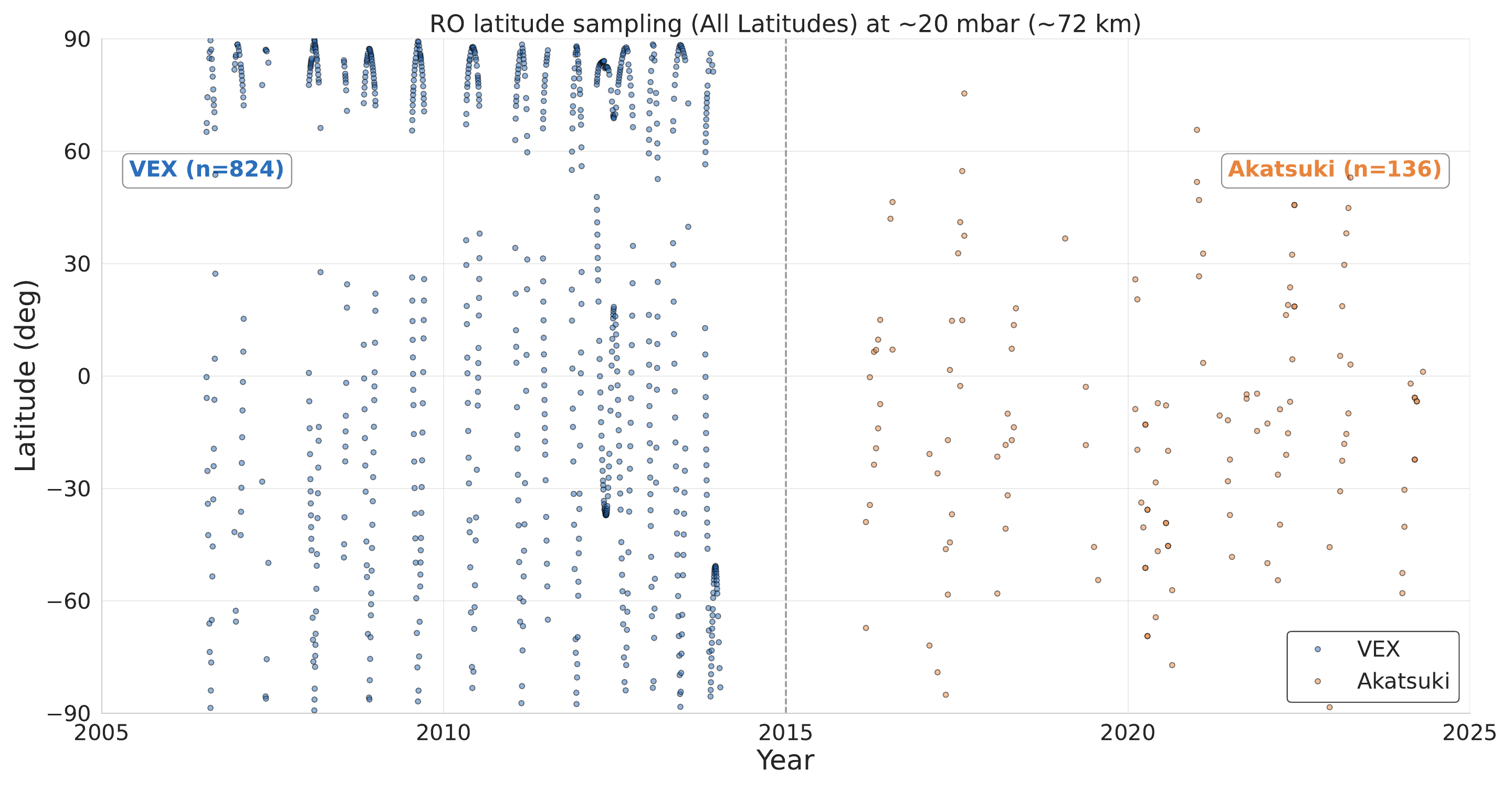}
    \caption{Latitudinal coverage of the RO experiments (VEX and Akatsuki) across the years from 2006-2024 at 20 mbar pressure ($\sim 72$ km altitude). The blue dots represent the VEX profiles and the orange ones are for Akatsuki. The dashed vertical line at 2015 roughly separates the missions into two regions year-wise, with no overlap between them. VEX has 824 profiles while Akatsuki provides 136 profiles.}
    \label{fig:1_lat_cov}
\end{figure*}

\begin{figure*} [tbh]
    \centering
    \includegraphics[width=1.0\linewidth]{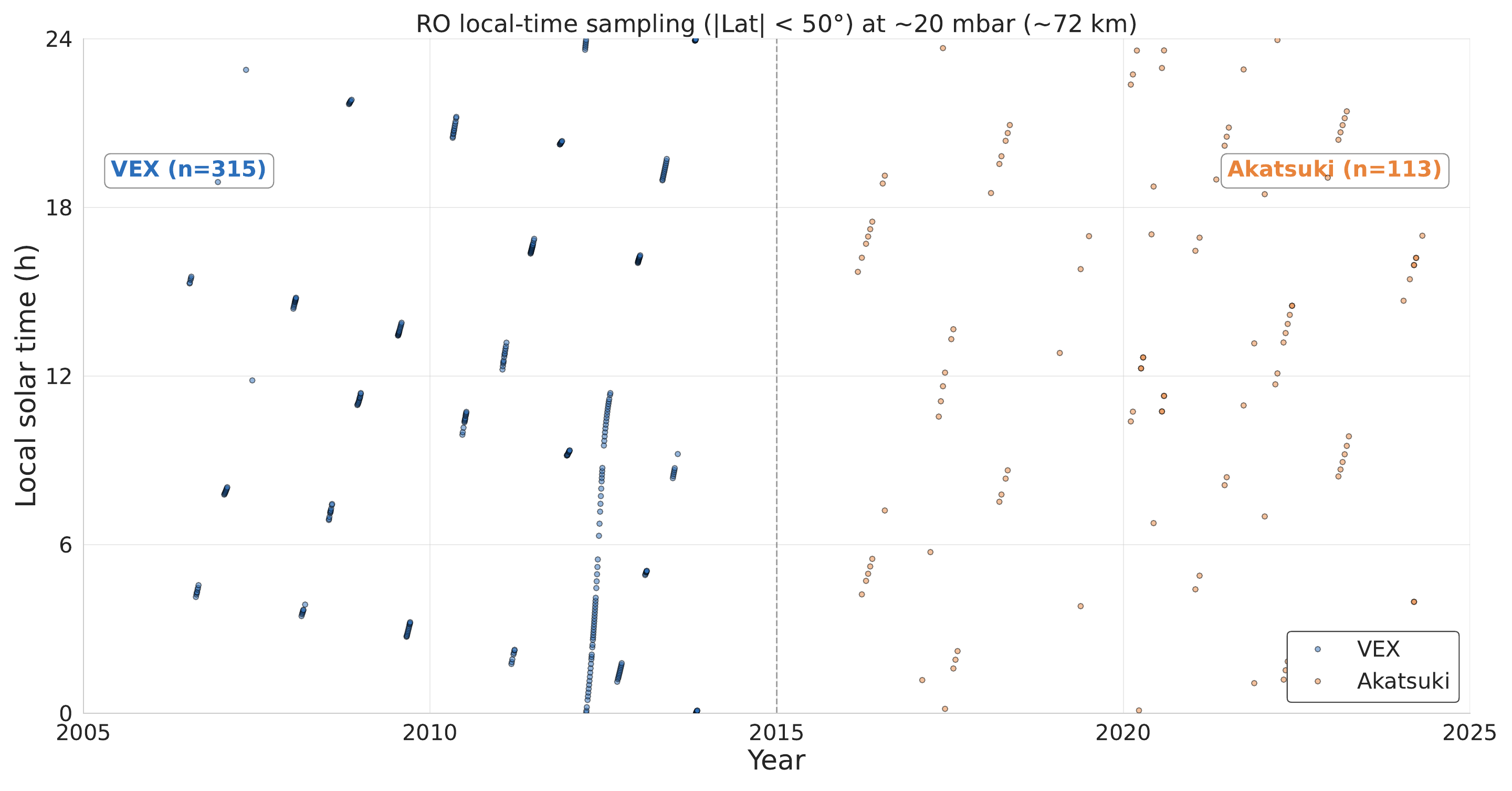}
    \caption{Local time coverage of the RO experiments (VEX (blue points) and Akatsuki (orange points)) across the years from 2006-2024 for the Region A latitudes. VEX has 315 profiles while Akatsuki has 113 profiles.}
    \label{fig:2_lst_cov}
\end{figure*}

The dense atmosphere of Venus below 45 km produces large fluctuations in the received RO signal, leading to larger uncertainty in the derived temperatures, and the boundary condition at 100 km also increases the uncertainty above 85 km \citep{Tellmann_2009}. Due to these factors, data from only 45-80 km altitudes are considered in this work. A total of 824 VeRa profiles and 136 Akatsuki RS profiles are included (taking both ingress and egress phases of the RO experiments) in this study. While VeRa predominantly sounds the northern high latitudes, Akatsuki mostly covers the low-latitude regions (Figure \ref{fig:1_lat_cov}). The local time RO sampling across the years for the low- to mid latitudes is also given in Figure \ref{fig:2_lst_cov}.

\subsection{Venus Climate Database}
\label{VCD_data}
The Venus Climate Database (VCD) is built upon the simulations of the Venusian atmosphere by the IPSL Planetary Climate Model (PCM) of LMD \citep{Lebonnois2010, Gilli2017, Martinez2023, lebonnois2024venus}. It provides mean values and statistics of atmospheric parameters such as temperature, density, winds, and also the abundance of certain atmospheric trace species such as SO$_2$, H$_2$SO$_4$ vapor, etc. It extends from the surface all the way to the thermosphere ($\sim$250 km). The IPSL PCM is essentially a general circulation model (GCM) that solves the primitive Navier-Stokes equations of atmospheric dynamics tuned to the Venusian conditions to simulate the atmospheric circulation and couples it with a full radiative transfer module that is able to generate the thermal structure self-consistently below 100 km \citep{GarateLopez2018, Scarica2019}. It also includes a fully coupled photochemical model and a simplified cloud model \citep{Stolzenbach2023}. The VCD gives high-resolution temporal outputs and is able to replicate the diurnal evolution of atmospheric phenomena such as thermal tides over a Venusian day \citep{Lai_2025}. VCD provides a horizontal resolution of 3.75° x 1.975° with an altitude resolution of ~1.9 km. It also allows the selection of different appropriate EUV scenarios and cloud albedo conditions to approximate the long-term variability in the thermal structure. VCD version 2.3 has been used in this work.

\subsection{Venus Global Reference Atmosphere(Venus-GRAM)}
\label{VIRA_data}

Venus-GRAM (VGRAM) is an engineering-focused atmospheric model developed by NASA to aid applications such as systems design, performance analysis, and entry, descent, and landing operations pertaining to Venus missions \citep{Justh_2006}. VGRAM queries the Venus International Reference Atmosphere (VIRA) to extract atmospheric parameters such as density, temperature, and winds all the way up to 250 km altitudes \citep{justh2017venus}. VIRA is a semi-empirical climatological model developed by integrating the PV Orbiter's OIR and ORO measurements, as well as the data from the Venera atmospheric probes \citep{Seiff_1985}. Below 100 km, which is our area of interest in this study, VIRA generates the profiles of the atmospheric parameters as a function of altitude and latitude. It provides the zonal mean, time-averaged atmospheric conditions that represent the climatological mean state of the Venusian atmosphere. 

\section{Results}
\label{Analysis}
\begin{figure*} [tbh]
    \centering
    \includegraphics[width=1.0\linewidth]{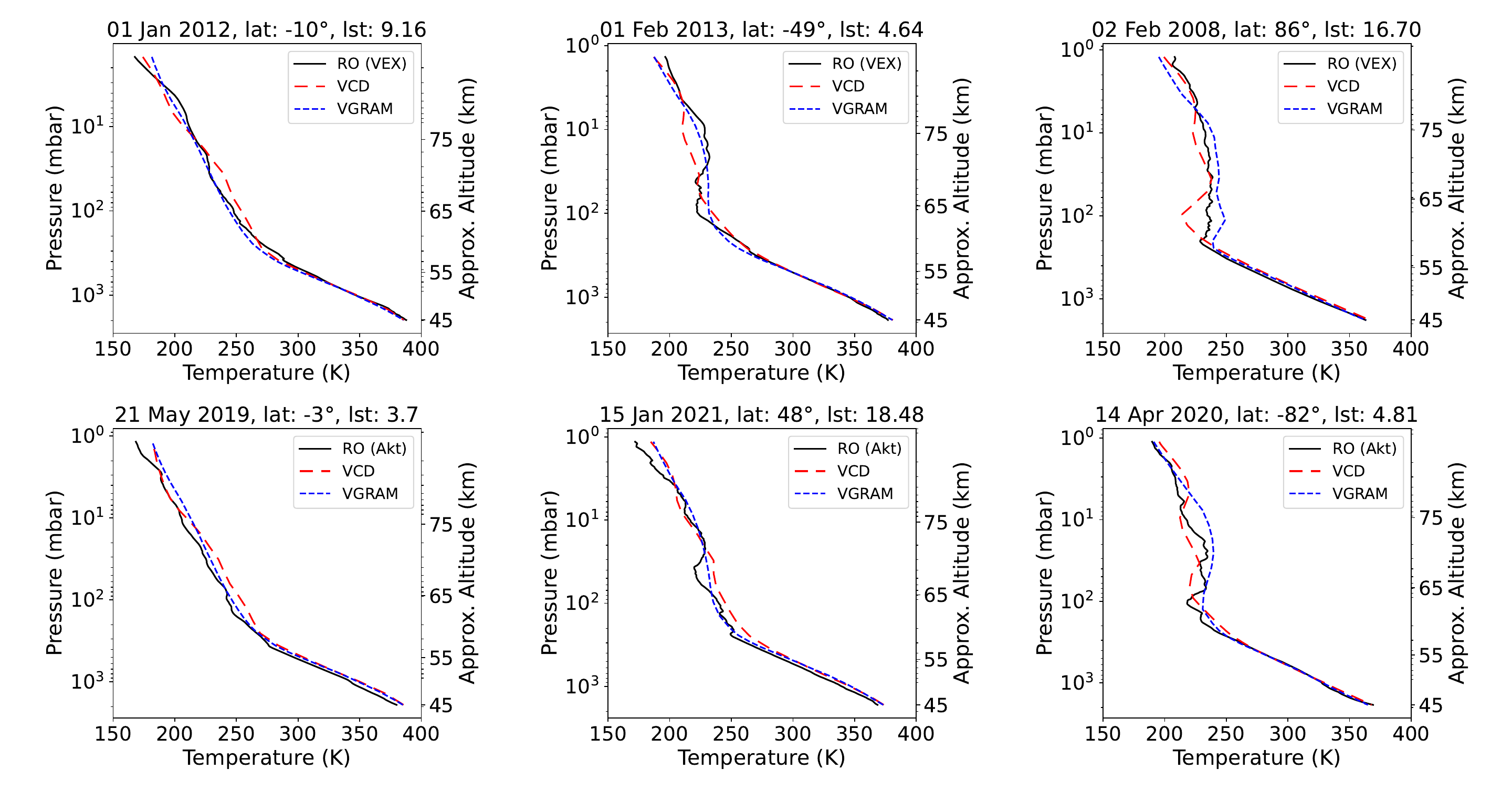}
    \caption{Temperature profiles in Venus for different latitude regions and local solar time plotted against pressure with the approximate altitude given in the secondary y-axis. The black profiles are derived from RO data, red profiles are generated using VCD and the blue profiles are from VGRAM. The top row corresponds to data from VEX while the bottom row contain data from the Akatsuki mission.}
    \label{fig:1a_single_profile}
\end{figure*}

In Figure \ref{fig:1a_single_profile}, sample RO temperature profiles from both VEX (top panels) and Akatsuki (bottom panels) missions from different latitude regions (low, mid, and high) and local solar times for six RO experiments are plotted along with the corresponding VGRAM and VCD temperature profiles. While there is good agreement between the observation and the models in the equatorial region (top left and bottom left panels), larger differences above $\sim60$ km emerge in the mid and high latitude regions. This difference becomes particularly pronounced in the high latitudes.

The VGRAM model was run with inputs corresponding to the latitude and altitude of each occultation point in VeRa and Akatsuki profiles, and the corresponding pressures and temperatures were derived. The pressures were then matched with the RO observations through interpolation, and the corresponding temperatures were used in the study. Since VCD allows pressure coordinates to be used directly as inputs, we used RO values of pressure, latitude, longitude, and local time of each occultation point, and the temperature profiles were generated. 

\begin{figure} [tbh]
    \centering
    \includegraphics[width=1.0\linewidth]{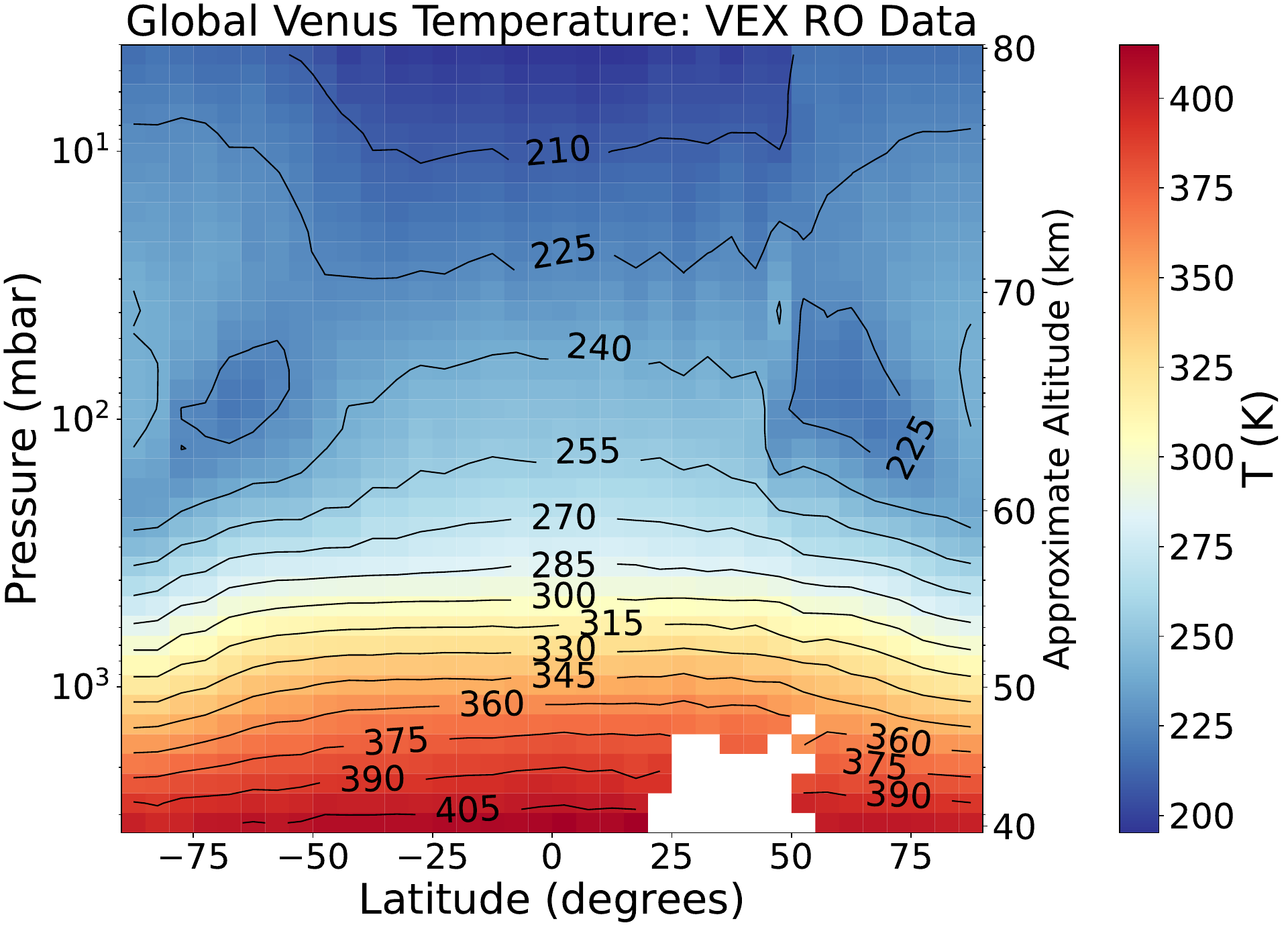}
    \caption{Global latitude vs pressure/altitude Temperature map of Venus with VEX RO data. The profiles are binned into latitude bins of $5^\circ$ and the pressure bins are equivalent to an approximate altitude of 1 km.}
    \label{fig:2a_VEX_global_map}
\end{figure}

\begin{figure*}[!t]
\centering

\begin{subfigure}{0.48\textwidth}
    \centering
    \includegraphics[width=\linewidth]{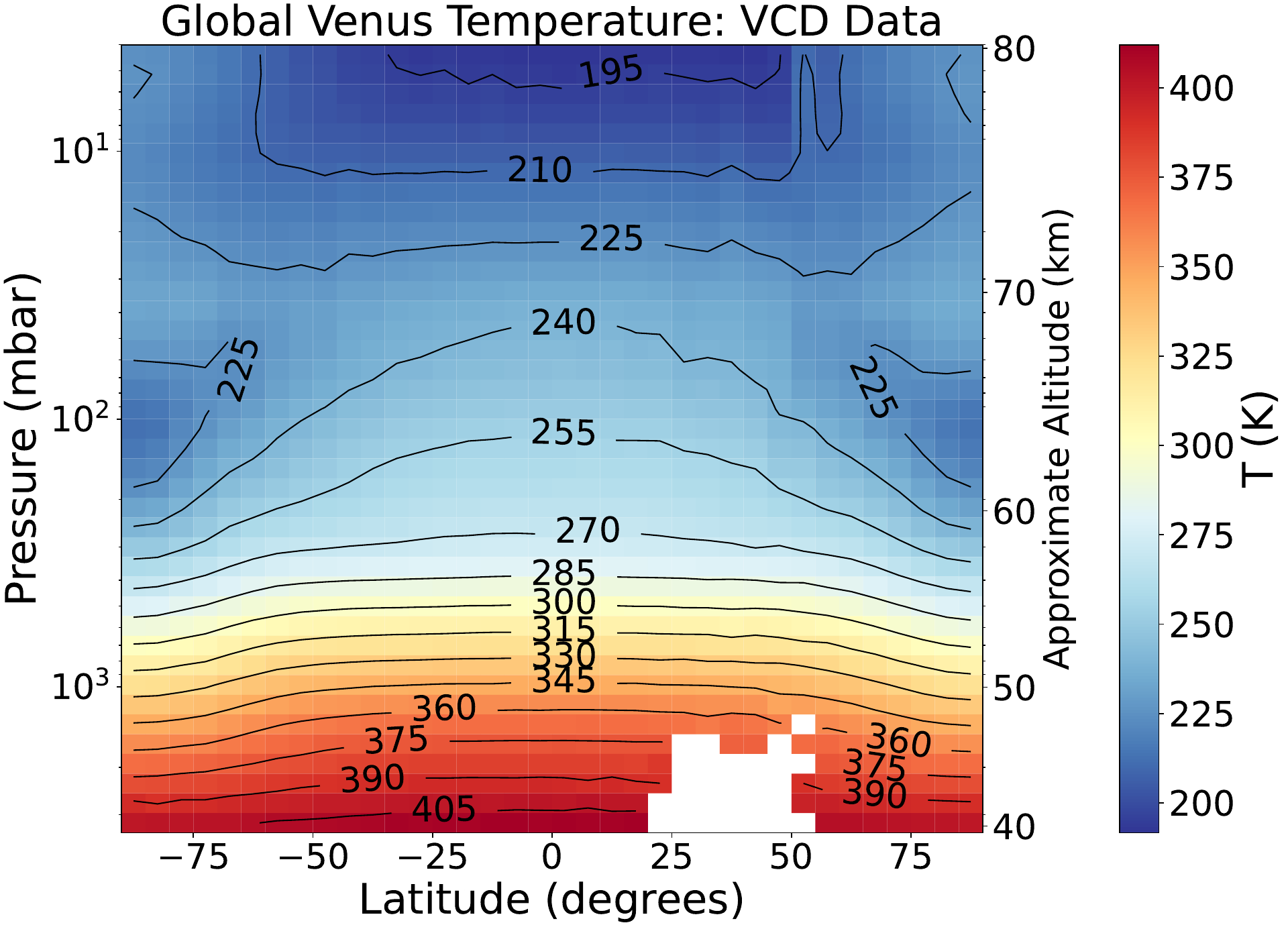}
    \caption{Global latitude vs pressure/altitude Temperature map of Venus with VCD. The inputs of latitude, longitude, local solar time and pressure are taken from RO samples.}
\end{subfigure}
\hfill
\begin{subfigure}{0.48\textwidth}
    \centering
    \includegraphics[width=\linewidth]{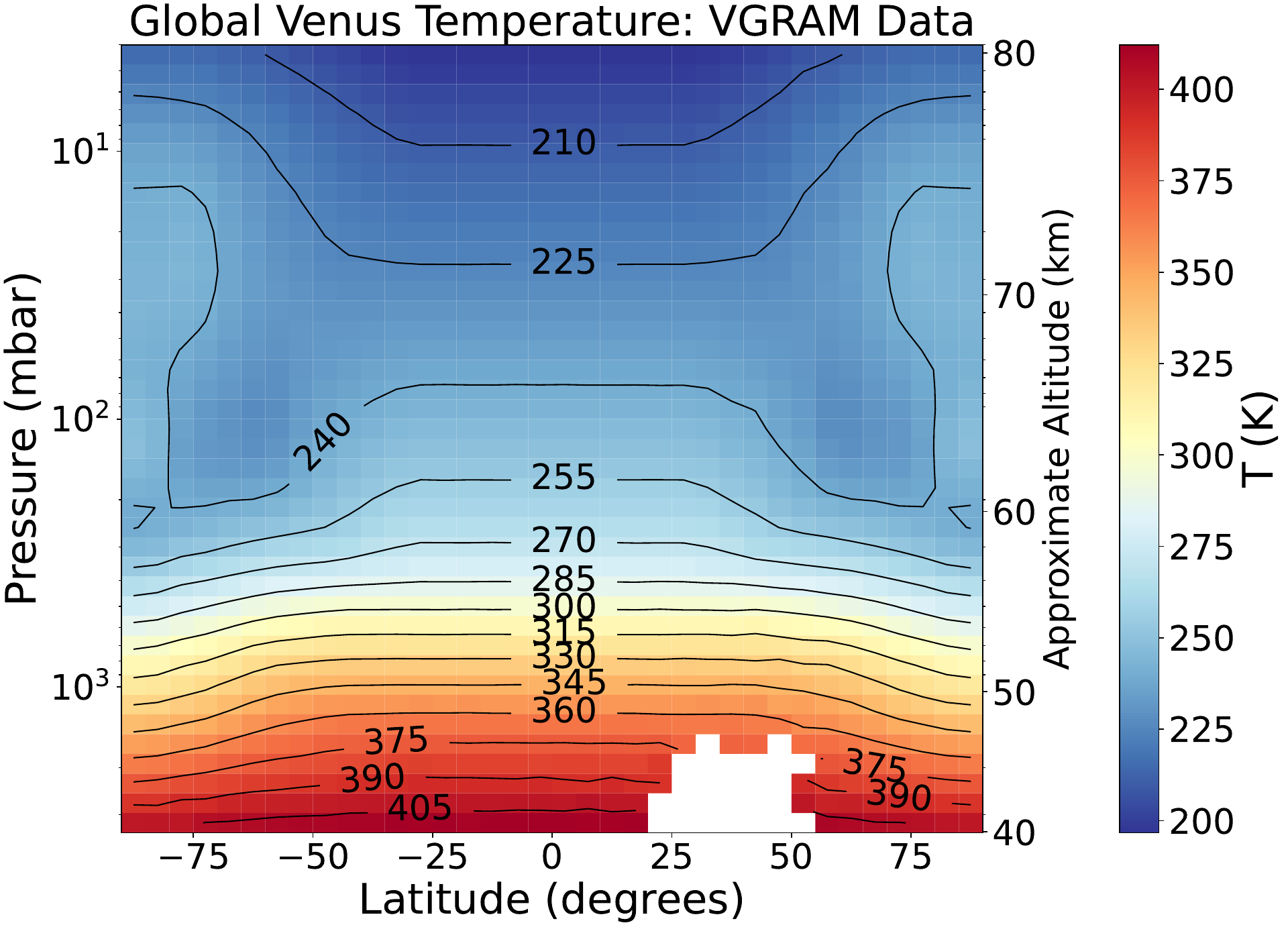}
    \caption{Global latitude vs pressure/altitude Temperature map of Venus with VGRAM. The inputs of latitude and altitude are taken from RO samples.}
\end{subfigure}

\vspace{0.5cm}

\begin{subfigure}{0.48\textwidth}
    \centering
    \includegraphics[width=\linewidth]{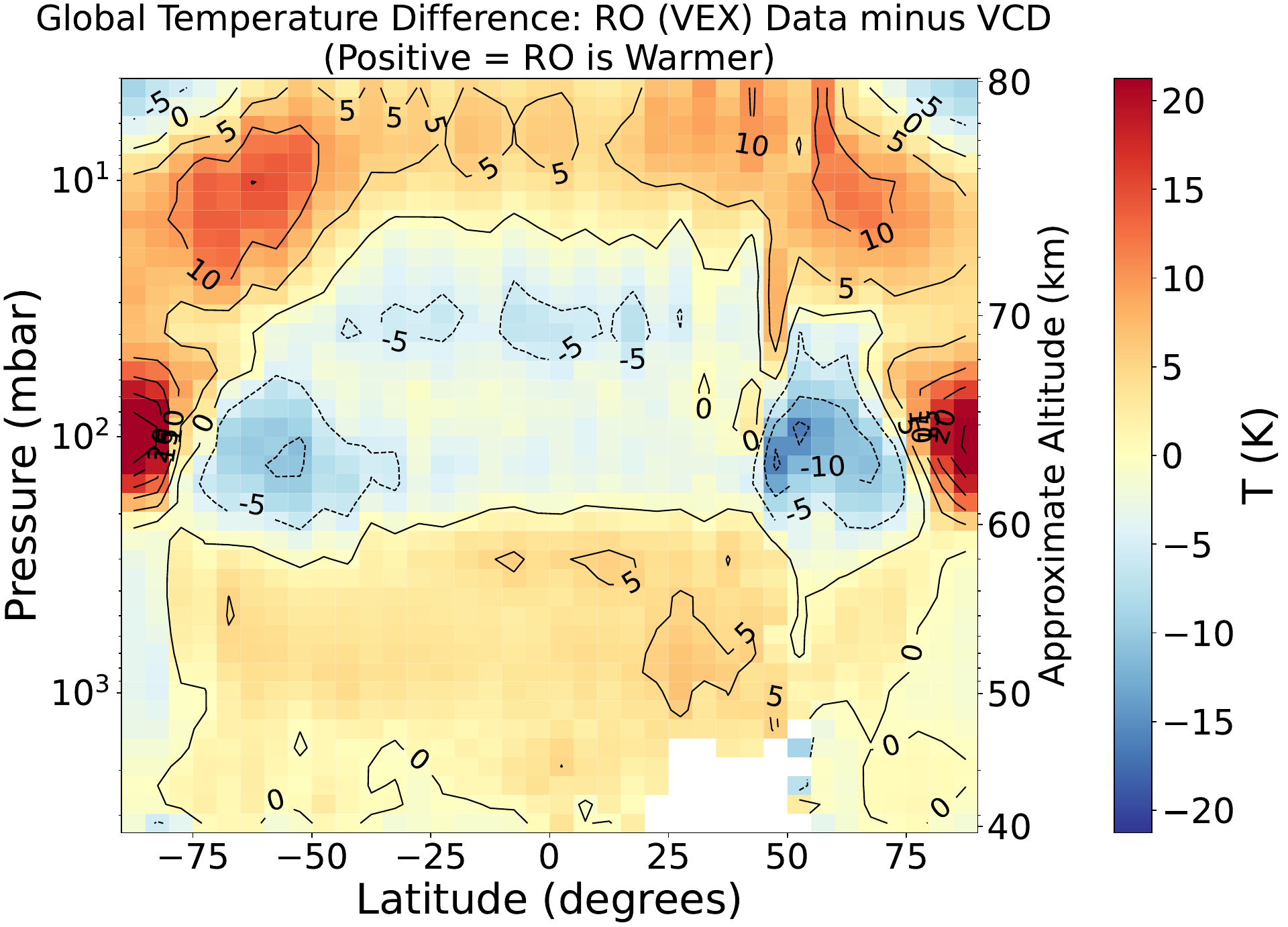}
    \caption{Global Temperature difference map of Venus between RO (VEX) and VCD. The difference is estimated at each point where RO data is present.}
\end{subfigure}
\hfill
\begin{subfigure}{0.48\textwidth}
    \centering
    \includegraphics[width=\linewidth]{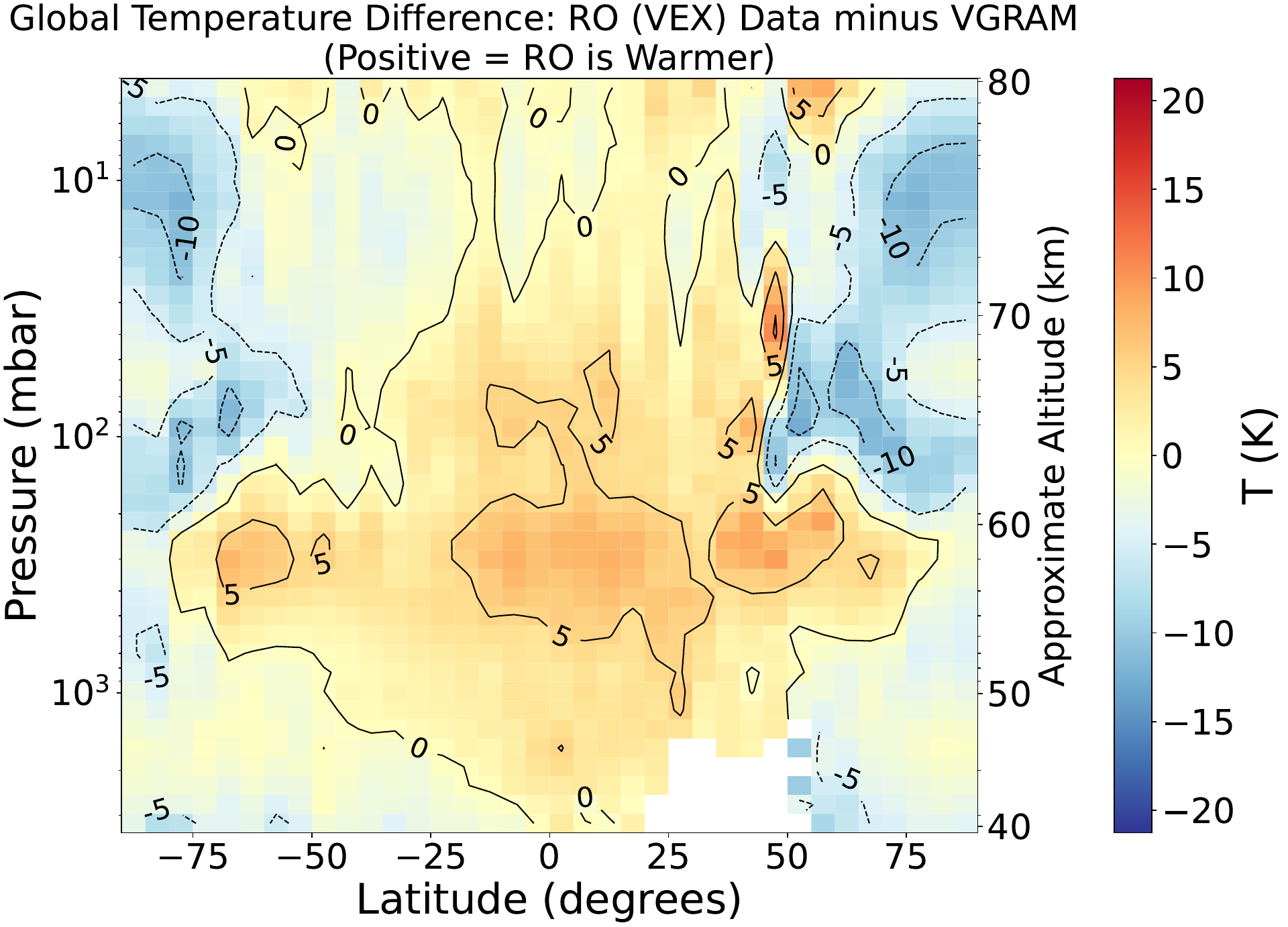}
    \caption{Global Temperature difference map of Venus between RO (VEX) and VGRAM. The difference is estimated at each point where RO data is present.}
\end{subfigure}

\caption{Global Temperature contour maps of Venus with VEX RO serving as the primary reference dataset. The data binning is the same as in Figure \ref{fig:2a_VEX_global_map}}
\label{fig:2bc_VCD_VIRA_global_map}
\end{figure*}

To get a clearer perspective about the overall thermal structure, the global latitude vs pressure contour map combining all the temperature profiles from the VeRa observations has been plotted in Figure \ref{fig:2a_VEX_global_map}. The corresponding VGRAM and VCD global maps are shown in Figures \ref{fig:2bc_VCD_VIRA_global_map} (top panels). The approximate altitudes corresponding to the given pressure levels are also shown. In the low latitudes, the temperature falls monotonically with increasing altitude and no temperature inversions are observed. This scenario changes drastically as we move towards the higher latitudes beyond $\pm$ 50$^\circ$ above $\sim$ 60 km altitudes, where the cold collar is first encountered. This localized cold patch, centered around 65$^\circ$ latitude, shows a difference of 10-15 K as compared to the low and high latitudes at the same height/pressure level. Upwards of 65 km, the temperatures are observed to increase as we move poleward. This feature is seen in both the northern and the southern hemispheres. A global map using Akatsuki RO profiles could not be constructed due to its insufficient coverage across the high and polar latitude regions of Venus (Figure \ref{fig:1_lat_cov}).

The VCD global temperature map, shown in Figure \ref{fig:2bc_VCD_VIRA_global_map} (top left panel), is able to reproduce the general thermal structure quite accurately. Differences, however, emerge in the higher latitudes, where the cold collar is seen to extend deep into the polar regions as well, a feature which is not observed in the RO measurements. Similarly, the VGRAM map (Figure \ref{fig:2bc_VCD_VIRA_global_map}, top right panel) also replicates the RO observations quite well, in line with our expectations since VIRA (primary source of VGRAM temperatures below 250 km) itself was developed largely from Pioneer Venus RO measurements. However, the VIRA map appears to be substantially smoother than the RO observations, which may contribute to some of the differences between the model and the observations. One possible reason for this smoother representation could be the relatively low spatial resolution of the Pioneer Venus OIR data \citep{Taylor1979}, which constitute one of the inputs to the model in this altitude range. Moreover, VGRAM is designed to provide climatological mean values, and thus deviations from real data are expected. This is the case with VCD as well.

To elucidate the variations between the models and the observations, the differences between the global maps are also plotted in Figures \ref{fig:2bc_VCD_VIRA_global_map} (bottom panels). In Figure \ref{fig:2bc_VCD_VIRA_global_map} bottom left panel, the VCD temperatures are subtracted from their corresponding VEX RO profiles. As can be seen from the figure, the differences in the lower altitudes (up to $\sim$ 60 km) of the low latitudes ($\pm$ 30$^\circ$) are quite small (typically less than 5K). However, this deviation of the models from the observation increases significantly in the cold collar regions and polewards of the collar across both hemispheres. The blue patch in the collar section (between $\pm$ 40$^\circ$ and 70$^\circ$ latitudes, and between $\sim$ 60-65 km altitudes) indicates that the model is overestimating as compared to RO, and in the polar regions, in the same 60-65 km altitude range, there is a significant underestimation (by more than 15 K). 

Above $\sim$70 km, large differences ($>$ 10K) occur between the model and RO, typically in the mid-high latitudes (beyond $\pm$ 50$^\circ$). Figure \ref{fig:2bc_VCD_VIRA_global_map} bottom right panel shows the contour plot of the difference between RO observations and the VGRAM temperatures. Here we find larger differences ($>$ 5K) in the low and mid latitudes till $\sim$ 60 km altitude than with VCD (Figure \ref{fig:2bc_VCD_VIRA_global_map} bottom left panel), as VGRAM underestimates the temperatures as compared to RO. In the high latitudes, in both hemispheres, there is a significant overestimation by VGRAM, leading to differences exceeding 10K between 60 km and 80 km altitudes. Such large variability ($>$ 10K) between RO and VIRA/VGRAM has not been reported earlier, and it highlights the need for updating the VGRAM taking into consideration our improved understanding of the Venusian atmosphere from both the VEX and Akatsuki missions.

\subsection{Long Term Trend Analysis}
\subsubsection{Trend with RO Observations}
\label{Trend_analysis}
\begin{figure*} [tbh]
    \centering
    \includegraphics[width=0.95\linewidth]{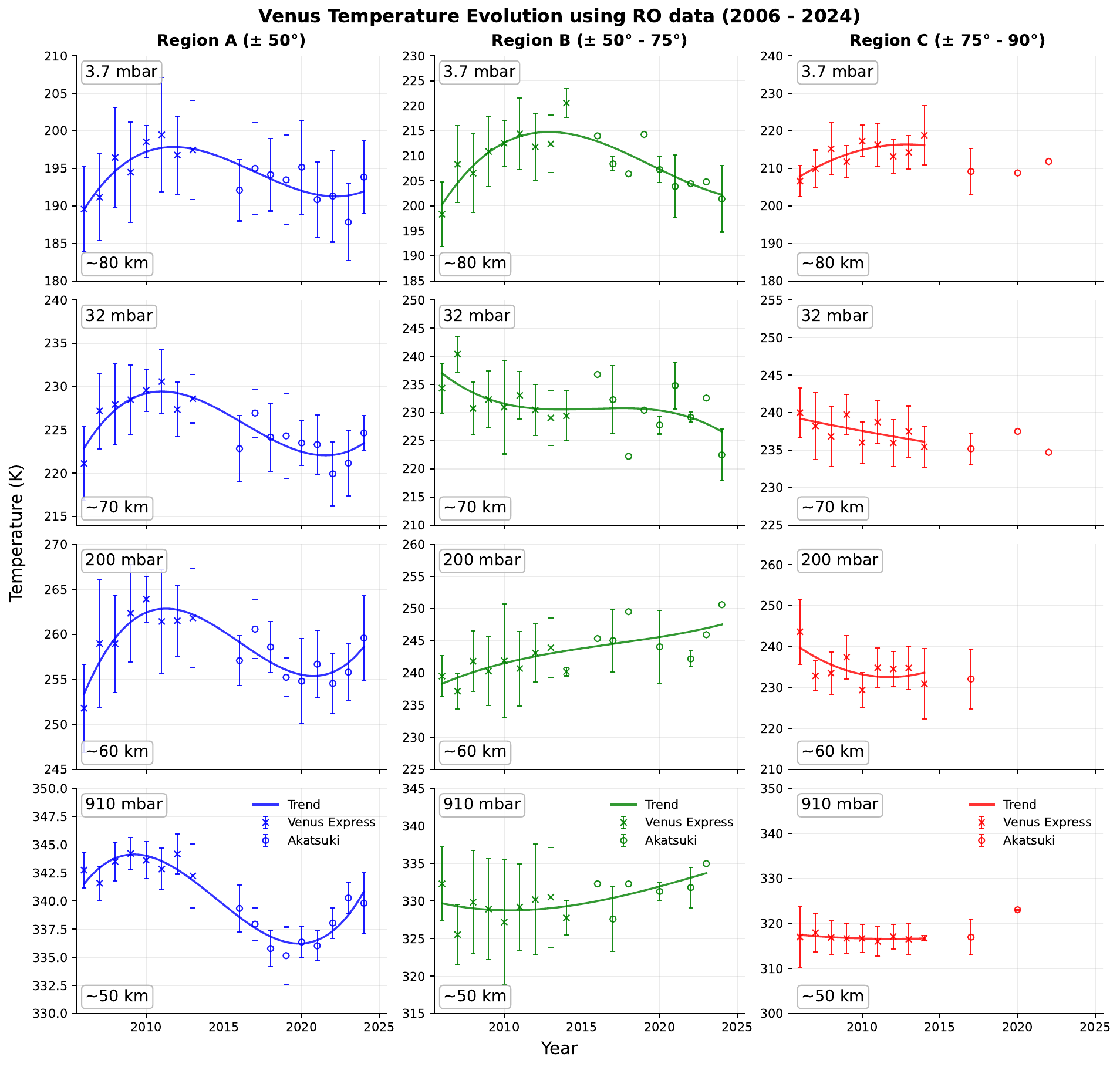}
    \caption{Long term evolution and trend analysis of Venus RO data from 2006-2024 between pressure levels of 910 - 3.7 mbar (approx. 50-80 km altitudes). The left panels show the Region A ($\pm50 ^\circ$) trends, the middle panels show the Region B (between $50^\circ$ and $75^\circ$ latitudes in both the hemispheres) trend while the right panels represent the Region C trend (beyond $\pm75 ^\circ$) in the data. The x marks are for the VEX points while o represent the Akatsuki data. A third order  polynomial trend lines are plotted for all the regions. The blue (Region A) and green (Region B) trend lines fit the entire 2006-2024 data, while the red (Region C) lines fit only the VEX data (2006-2014). The error bars represent 1 $\sigma$ deviation in the mean annual temperatures. The third-order polynomial fit shown only as a guide to the long-term variability. }
    \label{fig:4ab_long_term_trend_RO}
\end{figure*}

The evolution of the thermal structure over the two decades, from 2006-2024, is studied next. The data are divided into three latitude bands – (a) Region A ($\pm$ 50$^\circ$), (b) Region B ($\pm$ 50$^\circ$ - 75$^\circ$) and (c) Region C ($\pm$ 75$^\circ$ - 90$^\circ$). Region A corresponds to the low to mid latitudes, Region B covers the mid to high latitudes, and Region C comprises the polar latitudes in both hemispheres. The temperatures in each region are analyzed separately. The choice of the latitude bins was primarily driven by the highly dynamical nature of the cold collar regions, resulting in a potential latitudinal bias due to uneven sampling in the data. The effect of the bias is explored in more details in Section \ref{latbias}.  The data in each band are binned in pressure coordinates with the vertical spread of each pressure bin being approximately equivalent to 1 km in altitude. Then, the yearly mean profiles are calculated across all years along with the corresponding 1$\sigma$ error bars for different pressures (910 mbar, 200 mbar, 32 mbar and 3.7 mbar, for the approximate respective altitudes of 50 km, 60 km, 70 km and 80 km). Another possible bias in the data, due to uneven RO sampling in local times, can lead to large fluctuations in the annual mean temperatures. This can be due to diurnal and semidiurnal thermal tides in Venus and is corrected following \citet{Ando_2025}. The least squares method is used to fit the temperatures at each altitude with a sinusoidal function given by, 

\begin{equation}
T = T_0 + T_1 cos(2 \pi t /24 + d_1) + T_2 cos(2 \pi t/12 +d_2)
\end{equation}
where t is the local time (LT) in hours, $T_0$ is the temperature offset, $T_1$ and $T_2$ are the respective amplitudes of the diurnal and semidiurnal tide, while $d_1$ and $d_2$ are their respective phases at 0 LT. These thermal tide components are estimated separately for VEX and Akatsuki datasets and removed from the original RO data. A third-order polynomial fit is then applied to the annual mean temperature data to accurately capture the temporal trend at each pressure bin. This trend in the Region A latitudes (Figure \ref{fig:4ab_long_term_trend_RO} left panels) shows a potentially decadal-scale variability in the temperatures throughout the middle atmosphere (from approximately 50 km to 80 km). It should be noted that, similar to the findings in \citet{Ando_2025}, the thermal tide correction did not result in a significant change in the observed RO trend in our analysis either.

A similar long-term trend from 2006-2024 could not be drawn for the polar latitude (Region C) temperatures due to the significantly low number of profiles (Figure \ref{fig:1_lat_cov} and Table \ref{tab:year_count}) in the region from the Akatsuki mission because of its equatorial orbit (Figure \ref{fig:4ab_long_term_trend_RO} right panels). However, due to the high sampling of the latitudes between $\pm$ 75$^\circ$ and 90$^\circ$ by VeRa, we can trace the temporal evolution of the Venus middle atmosphere for the VEX years over this region. No consistent trend is seen over all the altitude bands in this region as it was observed for the Region A. While there is a decreasing trend at $\sim$ 50 km and 70 km altitudes, the temperatures are seen to increase near 80 km, and at $\sim$60 km, the mean temperatures are observed to stay almost constant. These inconsistent trends reflect the dynamical nature of the polar regions of the atmosphere, which are currently not well understood. The mid-to-high latitude belt (Region B), between $\pm$ 50$^\circ$ - 75$^\circ$, could not be accurately studied with these datasets due to the sparse sampling and the resulting latitudinal sampling bias in them, as highlighted in Section \ref{latbias}. Thermal tide correction was done for the VEX data across all the latitude bands, and only for Region A profiles of Akatsuki, due to inadequate sampling in the higher latitudes for the latter case (Table \ref{tab:year_count}). 

\begin{table}[!ht]
    \centering
    \caption{Sample counts across the years for each latitude band at 20 mbar pressure ($\sim$ 72 km altitude). 2006-2014 contains data from VEX, while Akatsuki coverage is from 2016 to 2024.}
    \label{tab:year_count}
    \renewcommand{\arraystretch}{1.3} 
    
    \begin{tabular}{crrrr} 
        \hline
        \textbf{Year} & \textbf{Region A} & \textbf{Region B} & \textbf{Region C} & \textbf{Total} \\ \hline
        2006 & 13 & 14 & 18 & 45 \\
        2007 & 11 &  2 & 17 & 30 \\
        2008 & 46 & 20 & 77 & 143 \\
        2009 & 28 & 15 & 31 & 74 \\
        2010 & 25 & 10 & 30 & 65 \\ 
        % \noalign{\vspace{0.5em}} 
        2011 & 48 & 34 & 49 & 131 \\
        2012 & 92 & 26 & 52 & 170 \\
        2013 & 52 & 66 & 44 & 162 \\
        2014 &  0 &  2 &  2 & 4 \\ \hline
        2015 &  -- &  -- &  -- & -- \\ \hline
        % \noalign{\vspace{0.5em}}
        2016 & 14 &  1 &  0 & 15 \\
        2017 & 13 &  3 &  3 & 19 \\
        2018 & 10 &  1 &  0 & 11 \\
        2019 &  4 &  1 &  0 & 5 \\
        2020 & 19 &  6 &  1 & 26 \\ 
        % \noalign{\vspace{0.5em}}
        2021 & 14 &  2 &  0 & 16 \\
        2022 & 18 &  1 &  1 & 20 \\
        2023 & 11 &  1 &  0 & 12 \\
        2024 & 10 &  2 &  0 & 12 \\ \hline
    \end{tabular}
\end{table}

A comprehensive study on the long-term evolution of the Venusian atmospheric thermal structure was similarly conducted by \citet{Ando_2025}. However, as stated earlier, their work mostly focused on the low-latitude regions, up to $\pm30 ^\circ$. Drawing from the earlier work of \citet{Lee_2019}, \citet{Ando_2025} was able to link the trend in the cloud top temperatures to the changing UV albedo, as a reduction in the albedo leads to more absorption of the incoming solar UV flux by the atmosphere, thereby leading to an increase in the temperature in the relevant altitudes. The trend in the lower altitudes (below 60 km) was not definitively attributed to any direct cause or mechanism. It is interesting to note that in our present study, the trend similar to the one shown in \citet{Ando_2025} is observed to extend to approximately $\pm50 ^\circ$ latitudes, signifying the spatial extent of the long-term thermal variability in the atmosphere. 

\subsubsection{Impact of latitudinal sampling bias in RO observations}
\label{latbias}
Due to the uneven sampling of the Venusian atmosphere during RO experiments, it becomes crucial to ascertain the reality of any observed atmospheric variability. Latitudinal bias in each year can affect long-term trends, especially in the highly dynamic cold-collar latitudes. In that case, either the bias has to be removed from RO observations to reveal the true variability, or the trend analysis itself cannot be done accurately. Ideally, we would like to have uniform RO sampling over all latitudes for all years of the experiment. However, this is not possible due to constraints set by spacecraft orbit and occultation geometry. To identify any potential bias in the data, we implemented a post-stratification algorithm to delineate true variability from sampling bias and combined it with a stratified bootstrap to calculate confidence intervals for the estimated bias and its temporal trend.

First, the climatological mean temperature $T_0$ is calculated at each $5^\circ$ latitude by $\sim$1 km altitude (in pressure coordinates) cell and subtracted from each RO observation. This difference, the temperature anomaly, is expressed as $T' = T - T_0$. The use of temperature anomalies allows us to remove the static background and makes it easier to delineate true variability from bias. Next, the anomalies are binned into strata of latitude bands (Regions A, B, and C) and pressure bins (between approximate altitudes of 50–80 km), and the analysis is conducted on one stratum at a time. A stratum is subdivided into cells of $5^\circ$ latitude and $\sim$1 km vertical coverage, and temperatures can vary within the stratum. Following this, one year’s data is taken at a time for each stratum, and two anomaly means - raw and matched - are calculated for each year. The raw mean represents the average anomaly combining all profiles from a given year in that stratum. It is a simple average, and the presence of any equatorward or poleward bias gets reflected in the annual mean value, averaged across the hemispheres. For example, in Region B, if 2010 has substantially more soundings in the $50^\circ\text{–}55^\circ$ latitude bin and 2011 has more observations in the $70^\circ\text{–}75^\circ$ latitude bin, it will show as a false trend in the data.

The matched mean estimates the year’s average anomaly within each cell separately and then combines the cell averages using weights taken from a pooled record, which contains the fraction of all profiles across all years that fall in each cell. For instance, assume that from 2006 to 2024 in Region C, the $75^\circ\text{–}80^\circ$ latitude bin at a given pressure was sampled 60\% of the time, the $80^\circ\text{–}85^\circ$ bin was sampled 25\% of the time, and the $85^\circ\text{–}90^\circ$ bin was sampled 15\% of the time across both hemispheres. To estimate the matched mean in Region C for the year 2009, the mean anomaly at each cell is estimated, and the weighted mean for the whole band is calculated using the weights 0.60, 0.25, and 0.15 corresponding to the fraction of sampled data for all years. The sampling bias for the year is then the difference between the raw mean and the matched mean.

\begin{figure*} [tbh]
    \centering
    \includegraphics[width=1.0\linewidth]{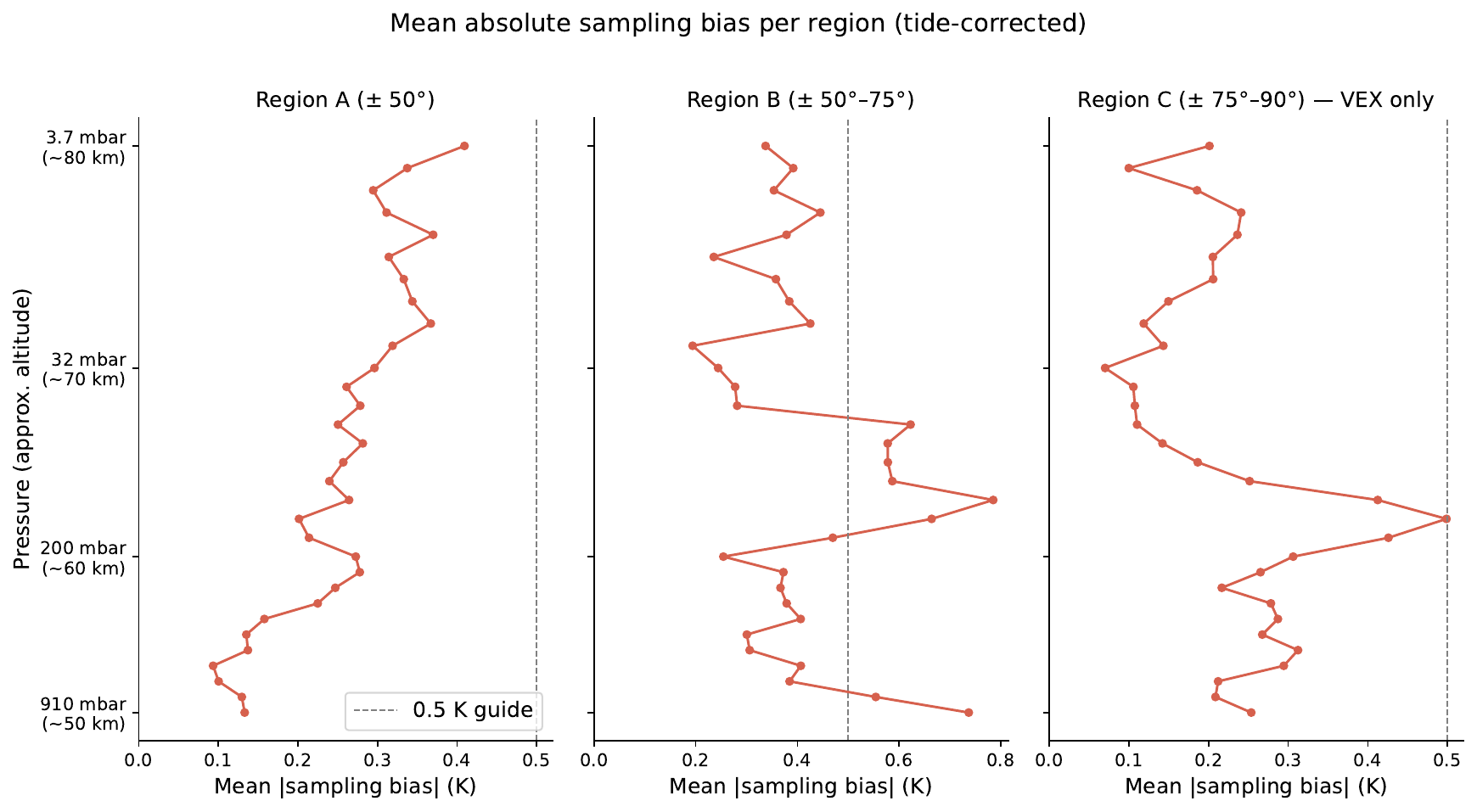}
    \caption{Mean Absolute Bias (MAB) in the RO temperature anomalies are plotted across the different latitude bands and approximate altitudes of 50-80 km. The dashed vertical lines represent the 0.5K threshold and any excursion beyond it indicates latitudinal sampling bias in the data.}
    \label{fig:MAB}
\end{figure*}

The limited number of profiles available to study sampling bias in the data can result in large uncertainty. To tackle this problem, the error is estimated using a resampling method. Within each cell, a subsample of the entire year’s profiles is drawn with replacement, and both the raw and matched means are estimated; this process is repeated two thousand times in our analysis. The spread of the two thousand recomputed values gives the uncertainty in the data. We compute the two means from the same subsample each time. To conclude whether the data is affected by sampling bias, we inspect (a) the mean absolute bias, (b) the trend in signed bias across the years, and (c) the reliability of coverage and the number of usable years in the data.The mean absolute bias (MAB) for each latitude region is plotted in Figure \ref{fig:MAB}. We have set a threshold of 0.5 K, beyond which the bias is taken to be significant. The MAB for Region C (Figure \ref{fig:MAB}, right panel) is estimated with VEX data only, due to the limited Akatsuki coverage in polar latitudes. We find that the latitude biases in Regions A and C fall within the 0.5 K threshold, while that in Region B (Figure \ref{fig:MAB}, middle panel) exceeds the threshold at $\sim$50 km and between 60 and 70 km altitudes.

\begin{figure*} [tbh]
    \centering
    \includegraphics[width=1.0\linewidth]{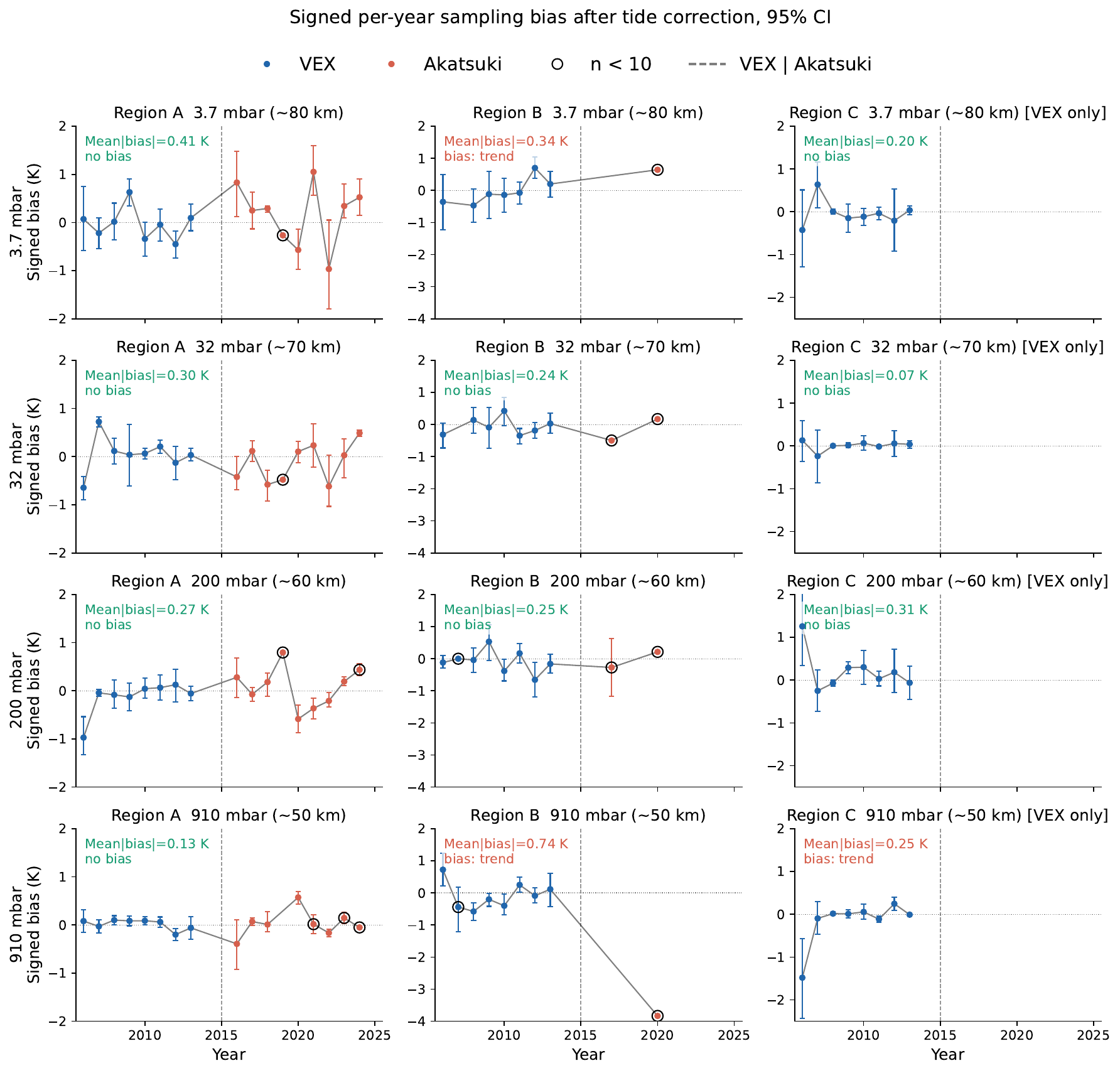}
    \caption{The temporal evolution of the annual signed sampling bias in the temperature anomalies is plotted for the three latitude bands at the 910 mbar ($\sim 50$ km), 200 mbar ($\sim 60$ km), 32 mbar ($\sim 70 $ km) and 3.7 mbar ($\sim 80$ km) pressure values. The blue points represent the VEX data and the orange points are the Akatsuki data. The vertical dashed line at year 2015 demarcates the separation between the two mission. The circle around a point signifies low sampling counts ($< 10$) at those years in the region. The error bars indicate the 95\% bootstrap confidence interval.}
    \label{fig:Signed_bias}
\end{figure*}

The signed per-year sampling bias along with the 95\% confidence interval (CI) is plotted in Figure \ref{fig:Signed_bias}. The blue points represent VEX data, while the orange points represent Akatsuki data. Circles around some of the points indicate low sampling for those years (counts $<$ 10). In Region A (Figure \ref{fig:Signed_bias}, left panels), the per-year bias is lower than 1 K, and no visible trend corresponding to the long-term variability in Figure \ref{fig:4ab_long_term_trend_RO} is observed in the data. The spread in bias points is noticeably smaller in the VEX years compared to the Akatsuki years because VEX has more data points. Thus, the analysis shows that the temporal variability in Region A cannot be explained by the latitudinal bias quantified using the above anomaly and post-stratification method, providing strong evidence for the robustness of the observed variability in the thermal structure.

For Region C (Figure \ref{fig:Signed_bias}, right panels), a large bias ($>$ 1 K) is observed in the 2006 data at lower altitudes (up to $\sim$60 km). However, after 2006, the spread in signed bias is very small, and we can conclude that the observed variation in Figure \ref{fig:4ab_long_term_trend_RO} for the region is not significantly impacted by latitudinal sampling bias as defined within the constraints of the present analysis. Nevertheless, the underlying physical cause of the variability in the region is not quantitatively known.

For Region B (Figure \ref{fig:Signed_bias}, middle panels), the number of reliable data points in the Akatsuki years is very low. We set a condition for the bootstrap analysis that subsampling of each cell in a stratum should only be performed if at least 3 data points are available for a given year. Akatsuki RO observations failed this criterion for 6 of the years in Region B. Additionally, even when the condition was met (in 2020, for example), the profile count for that year remained very low (fewer than 10). This makes any inference drawn from the Figure \ref{fig:4ab_long_term_trend_RO} Region B plots about the long-term evolution of the thermal structure in the region difficult to justify, and the trend is likely affected heavily by the lack of sufficient data. Interestingly, even in this region, the spread in the VEX bias is very small, and we can conclude that the VEX variability is mostly unaffected by any latitudinal sampling artifact.

\begin{figure*} [tbh]
    \centering
    \includegraphics[width=0.9\linewidth]{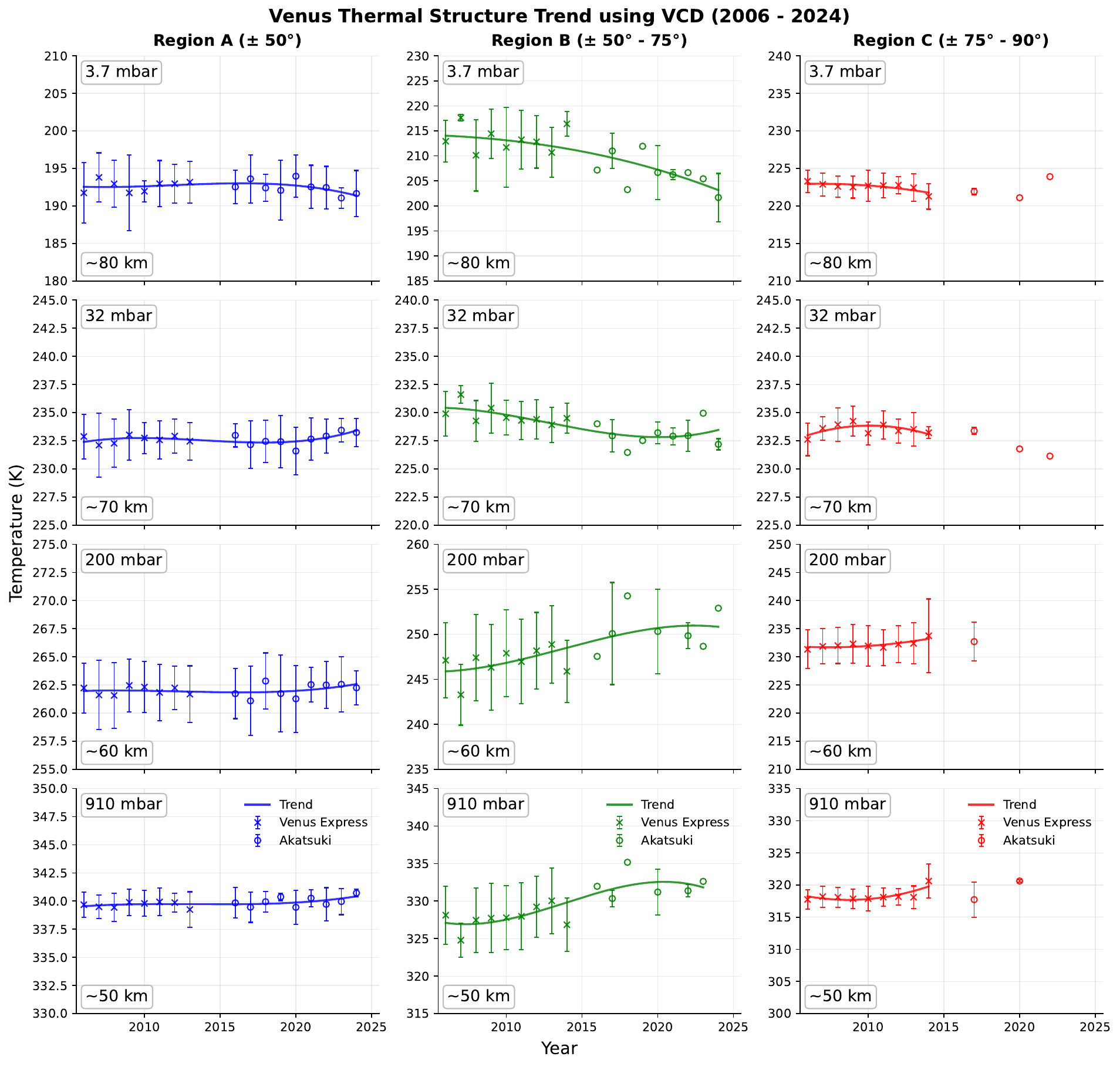}
    \caption{Trend in the Venus thermal structure of VCD from 2006-2024 between pressure levels of 910 - 3.7 mbar (approx. 50-80 km altitudes). The left panels show the Region A ($\pm50 ^\circ$) trends, the middle panels show the Region B (between $50^\circ$ and $75^\circ$ latitudes in both the hemispheres) trend while the right panels represent the Region C trend (beyond $\pm75 ^\circ$) in the data. The x marks are for the VCD outputs for the VEX years while o represent the Akatsuki years. Best fit polynomial trend lines are plotted for all the regions. The blue (Region A) and green (Region B) trend lines fit the entire 2006-2024 data, while the red (Region C) lines fit only the VEX years (2006-2014). The error bars represent 1 $\sigma$ deviation in the mean annual temperatures. The outputs are generated with constant cloud albedo setting of the \enquote{standard} condition in VCD.}
    \label{fig:6_VCD_trend}
\end{figure*}

\begin{figure*} [tbh]
    \centering
    \includegraphics[width=1.0\linewidth]{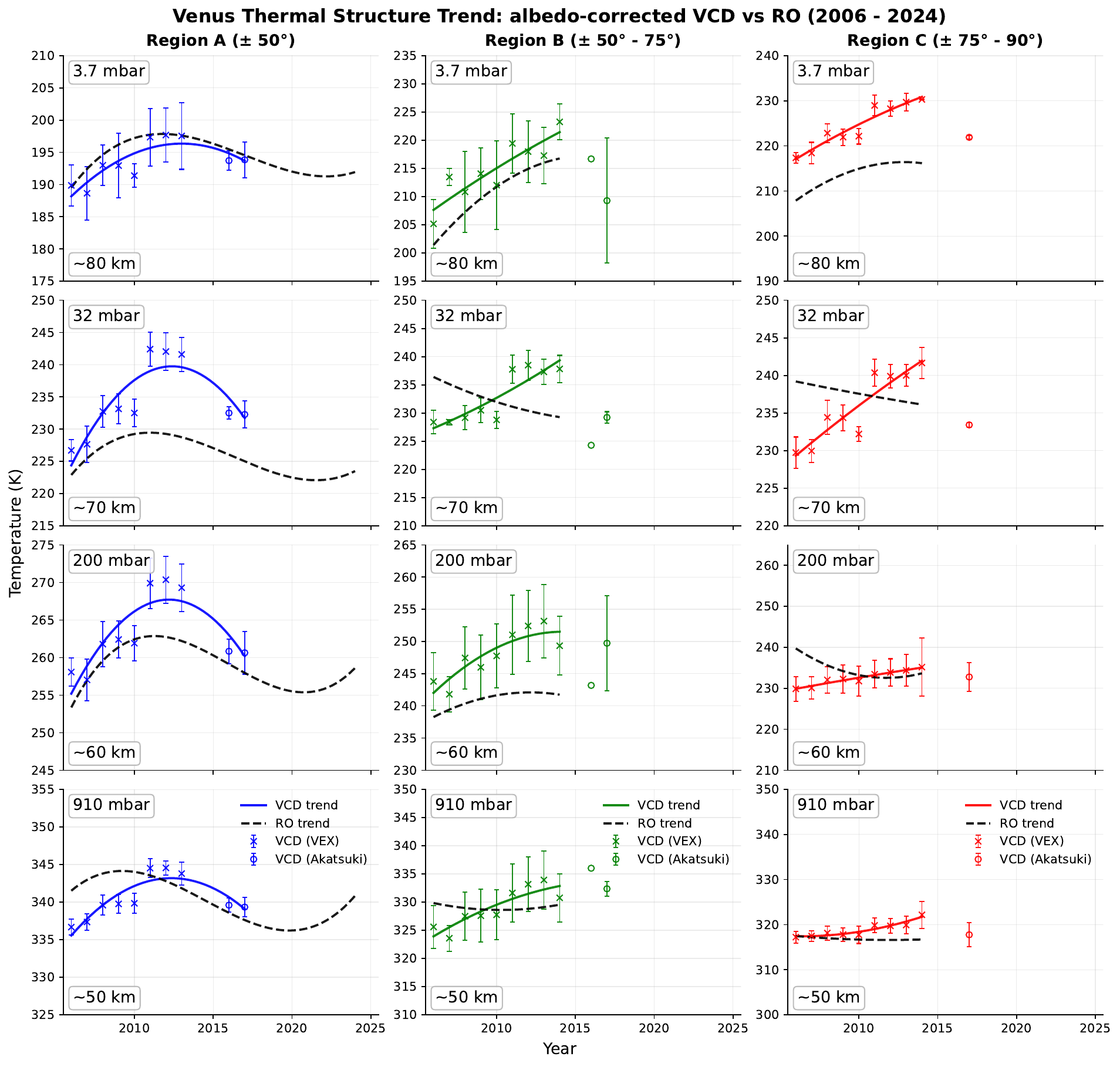}
    \caption{Trend in the Venus thermal structure of VCD from 2006-2024 between pressure levels of 910 - 3.7 mbar (approx. 50-80 km altitudes) with modified cloud albedo values. The blue (Region A) trend lines fit the 2006-2017 data, while the green (Region B) and red (Region C) lines fit only the VEX years (2006-2014). The black dashed lines represent the long term RO trends from Figure \ref{fig:4ab_long_term_trend_RO}. The cloud albedo conditions are set from \citet{Lee_2019}, where 2006 and 2007 are set as \enquote{high}, 2008-2010 is taken as \enquote{medium}, 2011-2014 correspond to \enquote{low} albedo values, and 2016-2017 again go back to \enquote{medium} values.}
    \label{fig:8_VCD_albedo_trend}
\end{figure*}

\subsection{Investigating the link between the cloud albedo and the cloud top temperatures using VCD}
\label{changing_albedo}

The VCD stores the climatological mean and statistics of the VPCM, which is run for a few Venusian days. It is incapable of generating any long-term evolution of the thermal structure from a physical point of view. However, when the latitudes, longitudes, and local solar times corresponding to the RO sampled points for each year (at a given pressure/altitude bin) are given as inputs to the VCD, we generate mean temperatures along with the standard deviation. Then the sampling data for every year from 2006 to 2024 are used to produce annual mean temperatures which can serve as proxy time series data, and we fit a polynomial similar to Figure \ref{fig:4ab_long_term_trend_RO} to show the trend in the temperatures across the years at different latitude regions and pressures (Figure \ref{fig:6_VCD_trend}). Although this trend does not represent the true evolution of the atmosphere in the GCM, it gives an approximate picture of the model outputs as the sampled RO regions change with time. The Region A trends (Figure \ref{fig:6_VCD_trend}, left panels) show little variability in the temperatures from 2006 to 2024. The Region C (right panels) trend line is drawn considering VEX data only, and here also a small variation of $\sim$2 K across the 8 years is observed, which is not consistent across all altitudes. Both these trends are significantly different from the RO variability observed in Figure \ref{fig:4ab_long_term_trend_RO} (left and right panels). Region B (middle panels) profiles show substantial fluctuations in the Akatsuki years, and it is difficult to comment on the thermal variations here. All the temperatures in Figure \ref{fig:6_VCD_trend} are generated with the constant \enquote{Standard} cloud albedo and EUV flux settings that are provided by the VCD.

As stated earlier, the characteristic variability in the Venusian atmospheric thermal structure across the years has been linked to the changing cloud-top albedo, which in turn primarily depends on the mysterious, still-unknown UV absorber \citep{Lee_2019, Ando_2025}. The albedo is seen to decrease from 2006 until 2012, when it reaches a minimum and then starts increasing again. The albedo data analyzed in \citet{Lee_2019} extends until 2017, providing a link between Venus Express and Akatsuki RO measurements. The trends seem to match for both the low and the high latitude regions, although the higher latitude albedo is $\sim$50\% greater than the low latitude one.

The public version 2.3 of the VCD gives the user the option to vary the cloud albedo conditions. This becomes an obvious parameter to tune and check how the global thermal structure varies according to the GCM with evolving background conditions. The VCD allows the cloud albedo condition to be set as one of three options— \enquote{low}, \enquote{medium}, and \enquote{high}. The three albedo settings correspond to three different solar heating rates used in the underlying GCM simulations, as described in the VCD documentation (ref: \url{http://www-venus.lmd.jussieu.fr/}): (a) The Standard (Medium) Albedo uses the standard heating rate profile based on \citet{Haus_2014}, (b) in the Low Albedo condition, the solar heating rate is estimated using the maximum ratio profile from \citet{Lee_2019}, scaled to a maximum of 50\%, and (c) the High Albedo setting takes the standard heating rate profile and decreases it by 30\%.

We have considered only the relevant VEX and Akatsuki years (2006–2017) for this comparison. The central idea behind carrying out this exercise is to check whether tuning the cloud albedo to crudely match the observations produces a better estimate of the model thermal structure instead of simply taking the default albedo condition and deriving the temperatures. The albedo conditions are set from \citet{Lee_2019}. 2006 and 2007 are considered to be \enquote{high}, 2008–2010 is taken as \enquote{medium}, 2011–2014 corresponds to the \enquote{low} albedo setting, and in 2016–2017 we return to the \enquote{medium} setting. This criterion is set for all latitude regions. The model is re-run and the temperature profiles are recalculated.

The long-term trend, shown in Figure \ref{fig:8_VCD_albedo_trend}, reveals an increasing trend in the VEX years across Region A, followed by a waning in the early Akatsuki years, beyond which cloud albedo data is not available. A similar increase in the VEX years until 2014 is observed in Regions B and C. The trend lines were not extended up to 2017 since not enough data points were available in 2016 and 2017 for thermal tide correction. The RO trends from Figure \ref{fig:4ab_long_term_trend_RO} are plotted alongside the VCD trend in Figure \ref{fig:8_VCD_albedo_trend} for comparison. We find that although the albedo-modulated VCD output is able to better mimic the RO observations, there is a distinct difference between the peaks of the VCD and RO trends, with the largest difference between the model and observations seen to be around 70 km altitude in Region A. It is quite possible that the availability of albedo values across all the years, until 2024, might have resulted in a more accurate fit. Additionally, more careful tuning of the cloud albedo conditions, which is not available in the current VCD version, might lead to a better estimation of the thermal structure.

\begin{figure} [tbh]
    \centering
    \includegraphics[width=1.0\linewidth]{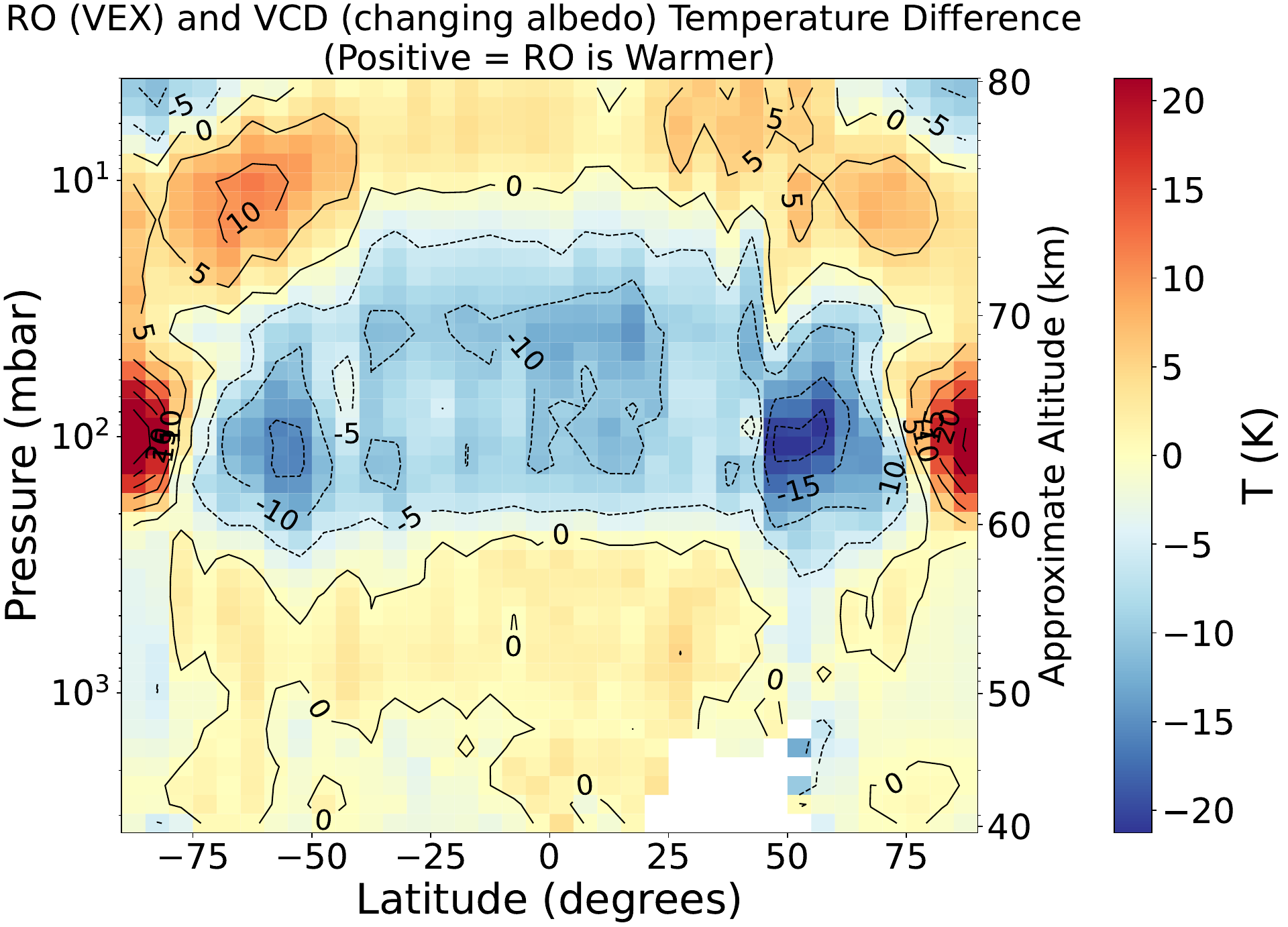}
    \caption{Global temperature difference map of Venus where the VCD output corresponding to VEX sampling is subtracted from the VEX RO data. The VCD temperatures are generated using varying cloud albedo conditions described in Section \ref{changing_albedo}.}
    \label{fig:6_VEX_VCD_changing_albedo}
\end{figure}

A contour map showing the difference between the RO observations and the albedo-modulated VCD temperature profiles is plotted in Figure \ref{fig:6_VEX_VCD_changing_albedo}. The difference over the low and mid latitudes up to 60 km altitude, which was already very small (Figure \ref{fig:2bc_VCD_VIRA_global_map}, bottom left panel), is seen to reduce further. However, above 60 km, there is a noticeable increase in the model temperatures which leads to an overestimation of more than 15 K compared to RO. This overestimation spreads up to $\sim$75 km in the low latitudes. Another point to note is that the difference between VCD and RO remains largely unaffected in the higher latitudes (poleward of $\pm60^\circ$). The increase in the VCD temperatures leads to a larger difference with RO in the northern cold collar region ($>$ 15 K instead of $>$ 10 K as seen in Figure \ref{fig:2bc_VCD_VIRA_global_map}, bottom left panel), while a smaller difference (of $\sim$5 K) is seen above 70 km around the $\pm70^\circ$ latitudes.

\begin{figure*} [tbh]
    \centering
    \includegraphics[width=1.0\linewidth]{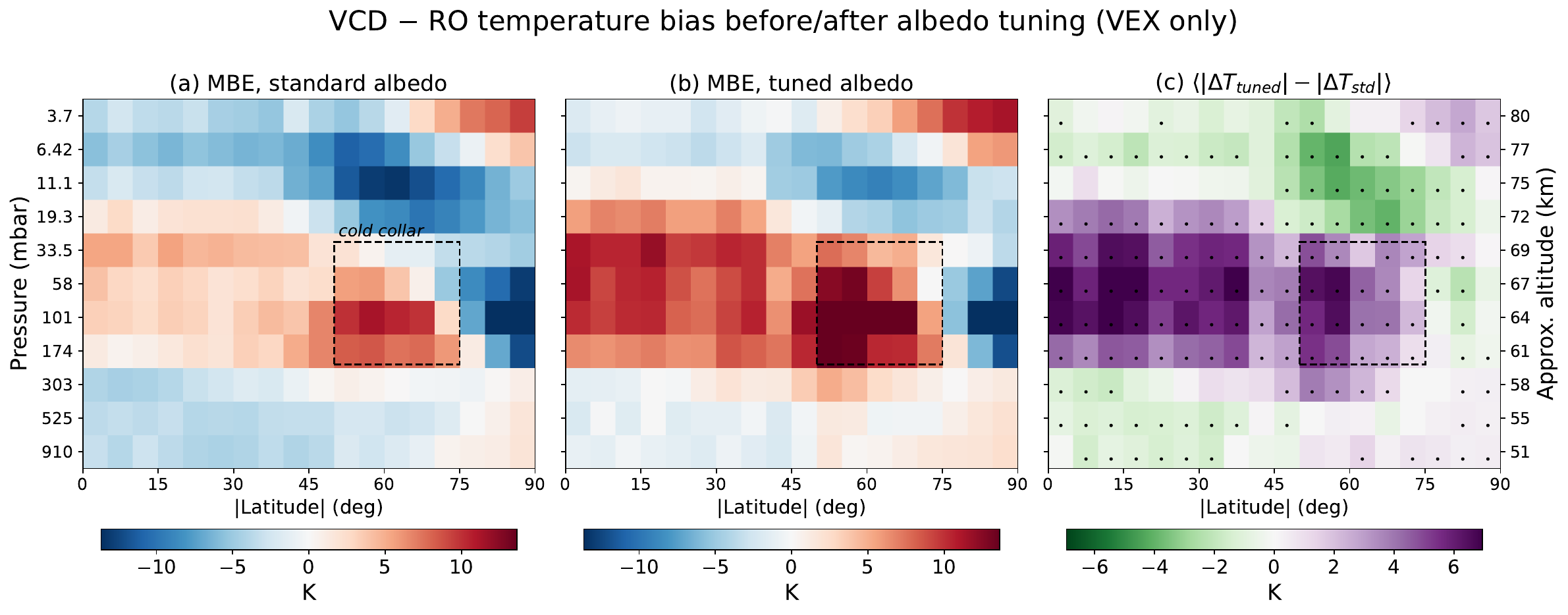}
    \caption{VCD - RO temperature bias before and after cloud-albedo tuning, plotted in latitude and pressure map. The latitude bins are hemispherically folded. (a) Mean Bias Error (MBE) for the standard albedo setting in VCD. (b) MBE for the albedo-tuned case in VCD. (c) Mean per-profile change in the absolute residual, $\left\langle \left| \Delta T_{\mathrm{tuned}} \right| - \left| \Delta T_{\mathrm{std}} \right| \right\rangle$ , where $\Delta T = T_{VCD} - T_{RO}$; the green color indicates that tuning the albedo has reduced the model's error while purple represents an increase in the error. The black dots in (c) mark cells where the 95\% bootstrap confidence interval on the mean paired difference (MBEs of tuned and standard albedos) excludes zero.}
    \label{fig:13_VCD_albedo_bias_heatmap}
\end{figure*}

\begin{figure*} [tbh]
    \centering
    \includegraphics[width=1.0\linewidth]{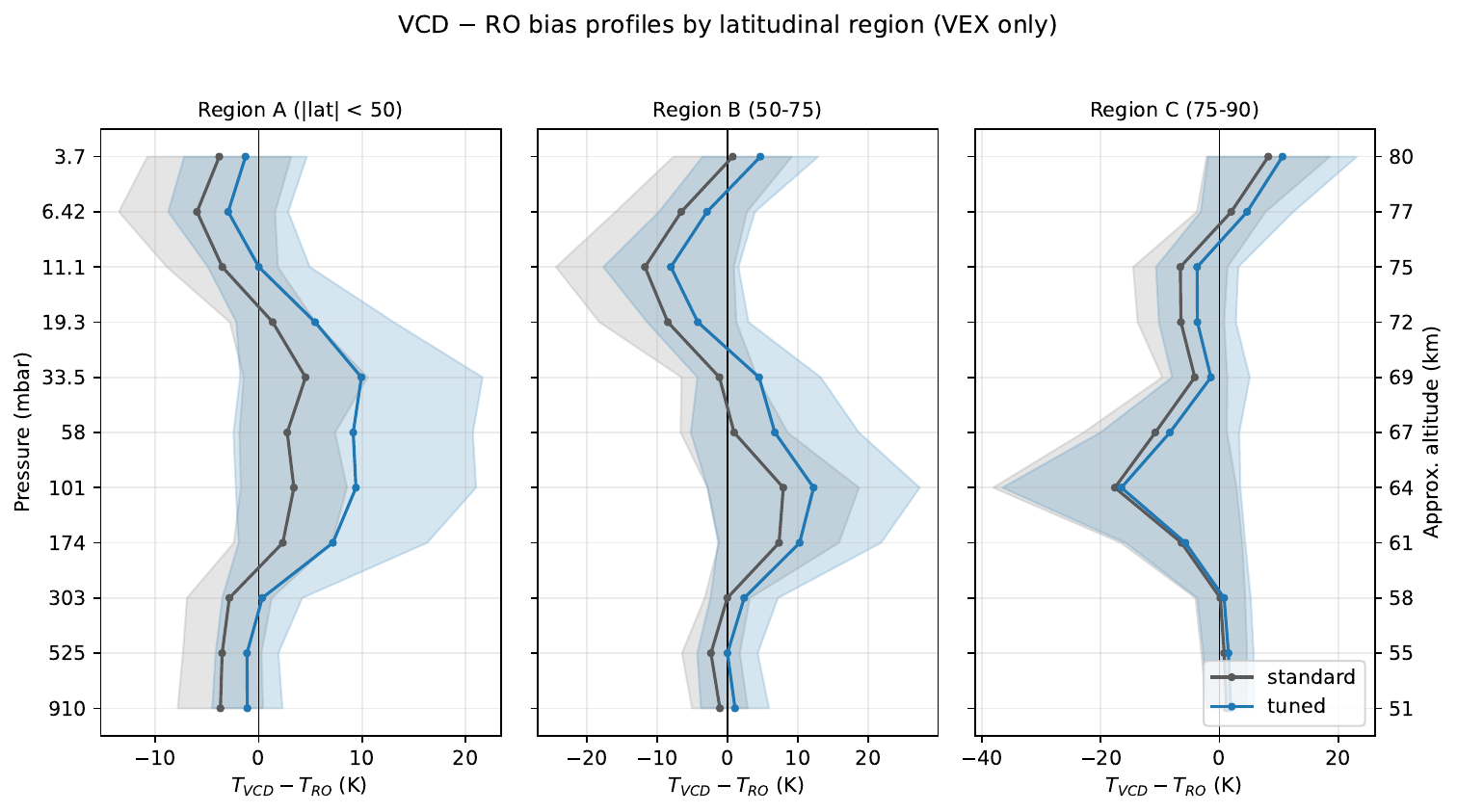}
    \caption{Vertical profiles of the VCD - RO temperature bias before and after the cloud albedo tuning, plotted for the three latitudinal regions A, B and C (defined in Section \ref{Trend_analysis}). Solid lines show the MBE; gray denotes the standard-albedo case while blue is for the tuned-albedo. The shaded bands span $\pm$RMSE about the mean. The vertical black line at 0K stands for the perfect agreement between model and observation.}
    \label{fig:14_VCD_albedo_MBE_RMS}
\end{figure*}

To better characterize the effect of the tuning of cloud albedo in VCD, we have also quantified the model biases before and after the tuning. This is shown in Figures \ref{fig:13_VCD_albedo_bias_heatmap} and \ref{fig:14_VCD_albedo_MBE_RMS}. The mean bias error (MBE) is calculated by subtracting the RO (VEX) temperatures from the VCD outputs at each point, separately for the standard albedo and the tuned albedo values, placed in isobaric pressure bins (between approximate altitudes of 50–80 km), averaged across the hemispheres, and further binned in $5^\circ$ latitude bins (Figure \ref{fig:13_VCD_albedo_bias_heatmap}, left and middle panels). The residual temperature is defined as $\Delta T = T_{\text{VCD}} - T_{\text{RO}}$. The mean per-profile change in the absolute residual temperature, $\delta_T = \left\langle \left\vert{} \Delta T_{\mathrm{tuned}} \right\vert{} - \left\vert{} \Delta T_{\mathrm{std}} \right\vert{} \right\rangle$, is plotted in Figure \ref{fig:13_VCD_albedo_bias_heatmap} (right panel) to show how the tuning improves or worsens the model predictions across the map. As already discussed earlier, there is significant overestimation by the model in the cold collar and polar latitudes around the cloud-top altitudes of 60–70 km in the standard albedo VCD run (Figure \ref{fig:13_VCD_albedo_bias_heatmap}, left panel), while the low-latitude differences are smaller. The MBE values for the tuned albedos (Figure \ref{fig:13_VCD_albedo_bias_heatmap}, middle panel) show an increase in contrast in the cold collar as well as the low-latitude regions between $\sim$70–80 km altitudes compared to the standard case. The $\delta_T$ values, plotted in Figure \ref{fig:13_VCD_albedo_bias_heatmap} (right panel), highlight this change quite clearly. The purple patches show that the tuning worsens the model predictions from the equator to $\sim$70$^\circ$ latitude. However, the polar latitude outputs remain largely unaffected. Additionally, below $\sim$60 km altitude and above $\sim$70 km altitude, the green bins indicate that the model estimates have improved after the tuning. This is most evident in the $45^\circ\text{–}70^\circ$ latitudes above 70 km height.

In Figure \ref{fig:14_VCD_albedo_MBE_RMS}, the MBE for standard and tuned albedo outputs along with the RMSE spread are shown across the different latitude bands and approximate altitudes of 50–80 km. We want to find if, by varying the albedo in VCD, the spread of the RMSE is tightened. The Region A latitudes (Figure \ref{fig:14_VCD_albedo_MBE_RMS}, left panel) see an increase in the RMSE spread between $\sim$58–72 km altitudes, and a decrease above and below this range. In the highly variable cold collar latitudes of Region B (Figure \ref{fig:14_VCD_albedo_MBE_RMS}, middle panel), the RMSE spread is again larger in the $\sim$58–70 km altitude range, but sees a tighter spread above this range. The Region C (Figure \ref{fig:14_VCD_albedo_MBE_RMS}, right panel) MBE and RMSE spread closely follow each other, with no significant difference. Thus, we conclude that the appropriate modulation of the cloud-top albedo is unable to sufficiently resolve the differences between VCD and RO observations.

\section{Discussion and Concluding remarks}
\label{Conclusion}

We have conducted a study on the thermal structure of the Venusian atmosphere below 100 km using RO observations as well as empirical and general circulation models. There is a growing need in the planetary science community for an updated and improved characterization of the Venusian thermal structure for upcoming missions to Venus, as well as to deepen our understanding of the evolution and sustenance of the dense and elusive atmosphere and the underlying physical processes.

More than 950 VEX and Akatsuki RO temperature profiles from 2006 to 2024 have been utilized in this work, the majority of them coming from VEX. On the modeling front, the popular empirical Venus-GRAM (VGRAM) and IPSL’s Planetary Climate Model (PCM), a general circulation model whose climatological mean output can be accessed through the Venus Climate Database (VCD) version 2.3, have been considered for comparative analysis.

This work builds on the results of \citet{Ando_2025} and indicates that the low-latitude variability in the thermal structure extends to mid-latitudes ($\le 50^\circ$). Furthermore, we have conducted an extensive analysis to assess the potential influence of latitudinal sampling bias on the observed RO trends. The results indicate that the observed temporal variability is not readily explained by the latitudinal sampling bias quantified using the adopted analysis, while sparse sampling limits the interpretation of trends at mid-to-high latitudes. The RO observations are then compared with the VGRAM and VCD models to highlight the existing discrepancies between models and observations.

Our study reveals that despite quantifiable progress over the decades in developing suitable empirical and general circulation models (GCMs), significant discrepancies still remain, highlighting inadequacies in our current understanding of the dynamical and chemical processes in Venus's atmosphere and its interactions with the surface as well as the Sun. This gap in our knowledge is largely due to the limited availability of data from sparse Venus missions, as well as the dense, opaque sulfuric acid clouds that prevent remote observations below cloud-top altitudes of $\sim$65 km.

The general thermal structure derived from both models matches well with VeRa observations at lower altitudes (up to $\sim$50 km). However, large deviations occur between the models and observations above 50 km, especially at mid- and high latitudes, within the cold collars and polar regions of both hemispheres in the 60--70 km altitude range. The cold collar generated in VCD extends into the polar regions, whereas RO observations find it to be confined within $\pm50^\circ$ and $75^\circ$ latitudes, thus contributing to the observed differences. VGRAM temperatures, queried from the VIRA model, display a very smoothed global map. A significant difference of more than 10 K in the thermal structure between VGRAM and VeRa is observed at polar latitudes ($T_{\text{VGRAM}} > T_{\text{VeRa}}$). As explained earlier, a possible reason for this difference could be the relatively coarse resolution of the Pioneer Venus OIR temperature retrievals, which constitute one of the inputs to VIRA. Because VGRAM represents the climatological mean state of the Venusian thermal structure, differences with day-to-day observations are expected, although previous reports have consistently shown discrepancies of less than 10 K.

Long-term analysis of the thermal structure shows possible decadal-scale variability at altitudes between 50 and 80 km in the low-to-mid latitude regions ($\pm50^\circ$). \citet{Ando_2025} have linked this trend at cloud-top altitudes to changing UV albedo as well as zonal wind speeds, suggesting potential coupling between dynamics, cloud microphysics, and photochemistry in this region. The reason for the occurrence of the trend at deeper altitudes remains unknown. A combination of post-stratification and bootstrap analysis on the sampled RO temperature anomaly data indicates that the observed long-term trends are not readily explained by the latitudinal sampling bias quantified using the adopted analysis across low-to-mid and polar latitudes, whereas sparse sampling in the mid-to-high latitudes by the Akatsuki spacecraft makes trend analysis inconclusive there.

The cloud albedo conditions and the solar EUV flux are inputs to VCD that were altered to test the dependence of the thermal structure on these parameters. It is seen that VCD, when run with evolving cloud albedo representing real observations, generates a trend similar to RO in the low-to-mid latitude region. However, the actual temperature differences between VEX data and VCD outputs increase above 60 km at lower latitudes. Model biases before and after tuning the albedo also reveal that the tuning produces spatially and vertically mixed results and fails to provide consistent agreement between VCD and RO observations. This test further highlights the limitations of the model in reproducing the observed long-term variability in the Venusian atmosphere and suggests that additional atmospheric processes may need to be considered.

As highlighted in other publications, planetary exploration will greatly benefit from upcoming Venus missions. Extended long-term monitoring with instruments onboard orbiters, in situ measurements, ground-based telescopic observations, and improved modeling efforts should lead to a more complete understanding of this enigmatic planet.

\section*{Acknowledgments}
We sincerely acknowledge the help and support of the ground station team at IDSN, India, and UDSC, Japan, to track the Akatsuki spacecraft. Special thanks to Umang Parikh, Anshuman Sharma, D32 Antenna, IDSN; and Himanshu Pandey, ISSDC, at Byalalu, India, for their proactive support. We acknowledge the support of Simi RS for archiving the data at SPL and making them available for analysis, of Arya Ashok, for helping us with some of the analysis part, and Keshav Aggarwal, for proof-reading and other useful suggestions during the revision of the work. We extend our gratitude to Dr. Siddharth Shankar Das and Nabarun Poddar at SPL for fruitful discussions on the sampling bias test. We also thank Prof. B. Hauesler, the principal investigator of Venus Express Radio Science payload, for archiving the frequency residual data sets. Special thanks to Ms. Ophrah Winfrey from Goa University, who initiated this work as a M.Sc. project at SPL.

\bibliographystyle{elsarticle-harv} 
\bibliography{paper2bib_new.bib}

\end{document}